\documentclass{aa}
\usepackage{txfonts}
\usepackage{graphicx}

\newcommand\phs{\phantom{$-$}}
\newcommand{\rb}[1]{\raisebox{1.5ex}[-1.5ex]{#1}}

\begin{document}

\title{
NGC~1261: An   $r$-process enhanced globular cluster  from the Gaia-Enceladus event
\thanks{This paper includes data gathered with the 6.5 meter Magellan Telescopes located at Las Campanas Observatory, Chile.}
}

\author{
Andreas J. Koch-Hansen\inst{1}
\and Camilla J. Hansen\inst{2}
\and Andrew McWilliam\inst{3}
  }
  
\authorrunning{A.J. Koch-Hansen, C.J. Hansen, A. McWilliam}
\titlerunning{$r$-process enhancement in NGC 1261}
\offprints{A.J. Koch-Hansen;  \email{andreas.koch@uni-heidelberg.de; hansen@mpia.de; 
andy@carnegiescience.edu}}

\institute{
Zentrum f\"ur Astronomie der Universit\"at Heidelberg, Astronomisches Rechen-Institut, M\"onchhofstr. 12, 69120 Heidelberg, Germany 
  \and Max-Planck Institut f\"ur Astronomie, K\"onigstuhl 17, 69117 Heidelberg, Germany
   \and Carnegie Observatories, 813 Santa Barbara St., Pasadena, CA 91101, USA
    }
\date{}
\abstract {
Our Milky Way (MW) has witnessed a series of major accretion events in the past. 
One of the later additions, the Gaia-Enceladus merger, has contributed
a considerable mass to the inner Galaxy, but also generously donated to the outer halo. 
So far, associations with present-day MW globular clusters (GCs) have been chiefly  
based on their kinematics and ages.
In this work, we present a chemical abundance study of the   
outer halo (R$_{\rm GC}$$\sim$18 kpc) GC NGC~1261, which has been suggested to be an accreted object based on its younger age.
We measured 31 species of 29 elements in two stars from high-resolution Magellan/MIKE spectra and find that
the cluster is moderately metal poor, at [Fe/H]=$-1.26$, with a low scatter of 0.02 dex. 
NGC~1261 is moderately $\alpha$-enhanced to the 0.3 dex level.
While from the small sample alone it is difficult to assert any abundance correlations, the light elements Na, O, Mg, and Al 
differ significantly between the two stars in contrast to the majority of other elements with smaller scatter; this argues in favor of 
multiple generations of stars coexisting in this GC. Intriguingly for its metallicity, NGC~1261 shows heavy element abundances that are consistent with 
$r$-process nucleosynthesis and we discuss their origin in various sites. 
In particular the Eu overabundance quantitatively suggests that one 
single $r$-process event, such as a 
neutron-star neutron-star merger or a rare kind of supernova, can 
be responsible for the stellar enhancement or even the enrichment of the cluster with the excess $r$ material. 
 Its heavy element pattern makes NGC 1261 resemble the moderately enhanced r-I stars that are
commonly found in the halo and have been detected in Gaia-Enceladus as well. 
Therefore, combining all kinematical, age, and chemical evidence we conclude that NGC~1261 is a chemically intriguing GC 
that was born in the Gaia-Enceladus galaxy and has been subsequently accreted into the MW halo.
}
\maketitle 
%
%
%
%
%
%
\section{Introduction}
Globular clusters (GCs) play an important role in building up the Galactic halo and bulge during an intense period of tidal disruptions
\citep{Gnedin1997,Odenkirchen2001,Martell2010,Schiavon2017,Koch2019CN,Hanke2020}. Equal importance is given to the accretion of the GC system of the Milky Way (MW) per se, when 
merging dwarf galaxies get stripped of their GCs during their gradual disruption \citep[e.g.,][]{SearleZinn1978,Sarajedini1995,Lee1999,Kruijssen2019,Malhan2019}.
Recently, several major merger events were identified in the  action space that had opened up with the powerful {Gaia} satellite \citep{GaiaDR2}
and/or via chemical tagging. Amongst 
these are Gaia-Enceladus \citep{Helmi2018GaiaEnceladus,Belokurov2018}, Sequoia \citep{Myeong2019}, and Heracles \citep{Horta2021}, all of which released stellar masses of a few 10$^{8}$ M$_{\odot}$
\citep[see also][]{Naidu2021}.
As one of the later additions (8--11 Gyr ago), Gaia-Enceladus has contributed
a considerable mass to the inner Galaxy, but also generously donated to the outer halo. In particular, \citet{Massari2019}  infer that 28 of the Galactic GCs 
are dynamically related to Gaia-Enceladus. 
So far, associations of such merger events with present-day MW GCs have been chiefly  
based on their kinematics and ages  \citep{Massari2019}, while chemical tagging of both components is only available for a few cases 
\citep[e.g.,][]{Koch2019Pal13,Koch2019Pal15,Horta2020}.

In this work, we focus on the outer halo GC NGC~1261, which has been long suggested to be an accreted system
based on a younger age of $\sim$10.3 Gyr \citep{Marin-Franch2009,Kravtsov2010}. 
While the distance values reported in the literature range from 15.5 kpc to 17.2 kpc \citep{Ferraro1999,Harris1996,Baumgardt2019,ArellanoFerro2019},
its location in the outer halo, at R$_{\rm GC}$$\sim$18 kpc, is undisputed.
So far, chemical abundance information for this cluster is sparse, and, to our knowledge,  
the only previous study is that by \citet{Filler2012}, who determined abundance ratios for 21 elements, but only reported on the GC mean and dispersion over their seven stars\footnote{Unfortunately, this work is only available as a AAS conference abstract and 
no details of the analysis are available.}. The light elements, in particular Na and O, measured by  \citet{Filler2012} show the strong variations that are commonly found  
across the Galactic GCs and that are strong indicators for the presence of multiple stellar populations in these systems \citep[e.g.,][]{Cohen1978,Carretta2009NaO,Bastian2018}.

In our present work, we further investigate the intrinsic properties of NGC~1261, but also attempt to place it in a Galactic framework by comparing our measured abundances
to various in situ and ex situ MW components.
A first step was taken by \citet{Filler2012}, who suggested that this {GC} shows lower $\alpha$-abundance ratios than otherwise found in systems at similar metallicity; 
our measurements, however, do not support these findings. 
It is interesting to note that \citet{Shipp2018} detected extratidal stars 
emanating stream-like from this GC, and an extended halo around NGC~1261 has been reported as well  \citep{Leon2000,Kuzma2018,Raso2020}. 
In the complex web of stellar streams, \citet{Shipp2018} place particular emphasis on the 
 EriPhe stellar overdensity \citep{Li2016}, which is suggested to be the remains of a tidally disrupted dwarf galaxy.
Moreover, a potential original association of EriPhe with the Phoenix stream  \citep{Balbinot2016} and 
NGC 1261 is highlighted, prompting the need for detailed chemical tagging, as we embark on the present study. 

This work is organized as follows: In Section 2 we introduce our target and observation strategies. Details on the chemical abundance analysis are given in Sect.~3, and the 
resulting abundance ratios are presented in Sect.~4. Next, Sect.~5 is dedicated to the $r$-process dominance in NGC~1261, before we discuss our findings in the context
of an accretion origin from Gaia-Enceladus in Sect.~6.
\section{Target selection, observations, and data reduction}
Two stars were selected from the catalog of \citet{Kravtsov2010}. Their choice was a trade off between avoiding the crowded, central regions of the GC, while 
opting for bright stars near the tip of the red giant branch (RGB). As a consequence of overall unfortunate weather conditions during the observing run, no more stars could
be targeted.
Our sample stars are shown in Fig.~1 together with the color-magnitude diagram (CMD) of \citet{Kravtsov2010} and an isochrone from the ``a 
 Bag of Stellar Tracks and Isochrones'' (BASTI) grid \citep{Hidalgo2018}, matching the age, metallicity, distance, and reddening of NGC~1261.
\begin{figure}[htb]
\centering
\includegraphics[width=1\hsize]{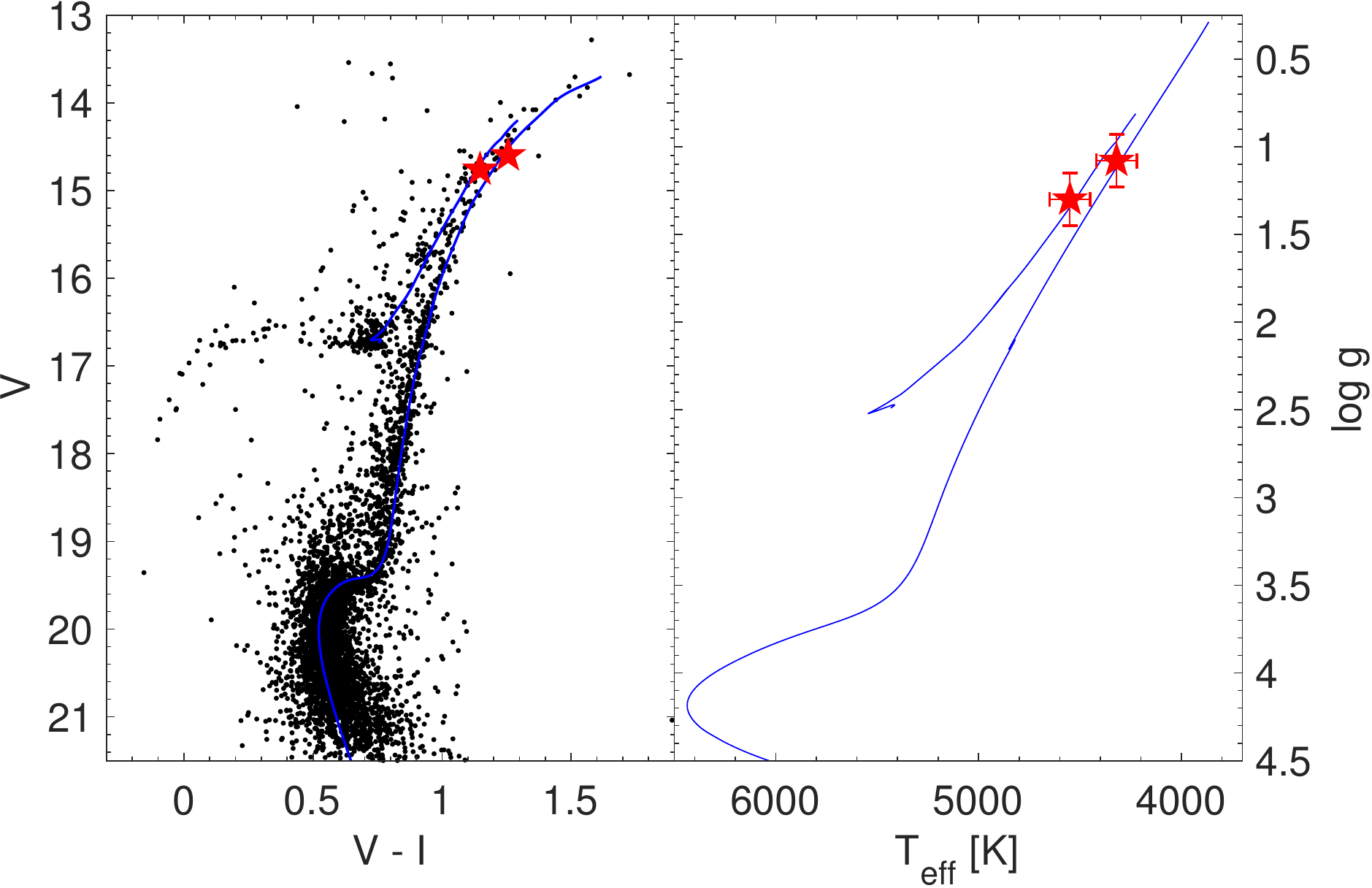}
\caption{CMD of NGC 1261 from the photometry of \citet[][left panel]{Kravtsov2010} and Kiel diagram (right panel). 
Our two targets are indicated as red symbols. Also shown is a 10.3 Gyr isochrone for the metallicity of our 
stars \citep{Hidalgo2018}.}
\end{figure}

Observations were carried out with the Magellan Inamori Kyocera Echelle (MIKE) spectrograph at 
the 6.5 m Magellan2/Clay Telescope at the Las Campanas Observatory, Chile during the night of Oct. 16, 2016. 
Under varying seeing conditions of $\sim$1.3--2.2$\arcsec$ we used a slit width of 0.5$\arcsec$ and binned the 
CCD pixels by 2$\times$1 in spatial and spectral direction. This resulted in  a resolving power of R$\sim$40,000. 
We make use of the red and blue arms of the instrument, covering the wavelength range 3340--9150\AA.  
The stars were exposed for 1 and 1.25 hours, which was split into four and three exposures, respectively, to facilitate cosmic ray removal. 

The raw data were reduced within the pipeline of Kelson (2000; 2003) and subjected to   
flat field division, order tracing from quartz lamp flats, and 
a wavelength calibration from Th-Ar lamp exposures. 
As a result, the  signal-to-noise ratios  (S/N) of the spectra reach  30--40 per pixel near the  peak on the order containing H$\alpha$, declining toward $\sim$15 
toward the blue at $\sim$4500\AA. 
In Table~1 we list the basic observational, positional, and photometric properties of the target stars. 
\begin{table*}[htb]
\caption{Properties of the targeted stars.}             
\centering          
\begin{tabular}{rccccccccccc}     
\hline\hline       
& $\alpha$  & $\delta$ & V & V$-$I  & t$_{\rm exp}$ & S/N\tablefootmark{b} & v$_{\rm HC}$ & T$_{\rm eff}$ & log\,$g$ & [Fe/H] & $\xi$  \\
\raisebox{1.5ex}[-1.5ex]{ID\tablefootmark{a}} &  (J2000.0) & (J2000.0)  & [mag] & [mag] & [s] & [pixel$^{-1}$] & [km\,s$^{-1}$] & [K] & [dex] & [dex]  & [km\,s$^{-1}$] \\
\hline
35  & 03:12:29.59 &  $-$55:11:30.19 & 14.597 & 1.256 & 4200 & 20/15/42/63 & 73.2 & 4320 & 1.08 & $-1.24$ & 2.00 \\
46 & 03:12:09.51  & $-$55:14:58.051 & 14.760 & 1.147 & 3600 & 18/12/33/50 & 69.9 & 4550 & 1.30 & $-1.27$ & 2.30 \\
\hline
\hline                  
\end{tabular}
\tablefoot{
\tablefoottext{a}{Identifications from the catalog of \citet{Kravtsov2010}.}
\tablefoottext{b}{Measured at 4500, 5200, 6600, and 8600 \AA, which coincides with the peaks of the  blaze function in the respective orders.}
}
\end{table*}

Radial velocities of the targets were determined by a cross-correlation of the red spectra
against a synthetic spectrum of an RGB star with stellar parameters representative of our targets. 
This yielded typical uncertainties of $\sim$0.1 km\,s$^{-1}$. 
Clustering around a mean heliocentric value of 71.5$\pm$1.2 km\,s$^{-1}$, these stars are confirmed to be members with NGC~1261, 
which has been recently reported to have a systemic mean heliocentric velocity of %
71.36$\pm$0.24 km\,s$^{-1}$ \citep{Baumgardt2019}.
The small sample size of only two stars; however, inhibits an in-depth study of the dynamic properties of the GC.
\section{Abundance analysis}
The chemical composition of the two stars was determined from a standard analysis based on equivalent width (EW) measurements and spectral synthesis. 
For the former, we fitted  Gaussian line profiles using the IRAF {splot} task.
We built our analysis on the line list from \citet[][and references therein]{Kacharov2013} for the moderately metal-rich GC M\,75. In order to obtain 
a fair balance of weak and strong lines, this was complemented with the Fe-line list of \citet{Ruchti2013}, which was optimized for solar-metallicity stars. 
Further additions in the spectral synthesis in the blue or noisier spectral ranges 
 are from 
\citet{Lawler2001,Lawler2006,Lawler2009},               
\citet{DenHartog2003}, and      
\citet{CJHansen2013}.           
Hyperfine splitting was included where appropriate (i.e., it was negligible for weaker lines) using data from 
\citet{DenHartog2011},  
\citet{CJHansen2013},   
\citet{Mashonkina2014},         
\citet{Lawler2015,Lawler2019}, and      
\citet{Shi2018}.                
In the actual abundance analysis we used the ATLAS grid of  plane-parallel,  72-layer, one-dimensional, line-blanketed Kurucz models without convective 
overshoot. Moreover, the $\alpha$-enhanced opacity distribution functions AODFNEW of \citep{CastelliKurucz2003} were adopted. 
We further adhered to  local thermodynamic equilibrium (LTE) for all species. 
All computations employed the stellar abundance code MOOG \citep[][version 2014]{Sneden1973}.
\subsection{Stellar parameters}
To first order, we estimated the effective temperatures based on the V and I photometry of \citet{Kravtsov2010} combined with 
the K magnitudes from 2MASS, 
and we adopted a reddening of 0.012 \citep{Schlafly2011}. The metallicity to be adjusted in the calibrations of T$_{\rm eff}$ with 
V$-$I and V$-$K colors by \citet{RamirezMelendez2005} 
was initially set at the  value reported by \citet[][2010 version]{Harris1996}. The dispersion between results from the two color indices
lies at 75 K on average. 

Using those T$_{\rm eff}$ values, photometric gravities were computed. To this end, we adopted the latest distance to NGC 1261 of 
15.5 kpc based on Gaia DR2 \citep{GaiaDR2,Baumgardt2019}, but we note that the range of distances reported for this GC  includes farther values
out to 17.2 kpc \citep{ArellanoFerro2019}. Another source of uncertainty is the mass of the stars, which we took as 0.8 M$_{\odot}$.
However, noting that star 46 may be on the asymptotic giant branch (AGB; see Fig.~1), a lower mass of $\sim$0.6 M$_{\odot}$ is also 
feasible. Combined, all these error sources results in an overall uncertainty in the surface gravity of 0.15 dex.  

In the next instance, the parameters were refined based on the balances of the iron abundance from neutral lines with excitation potential to settle T$_{\rm eff}$, and 
with reduced width so as to fix the microturbulence $\xi$.
The photometric and spectroscopic values are identical for star 35, while the latter is warmer by 110 K for star 46; we adopt the 
spectroscopic T$_{\rm eff}$ in the following.
As an estimate for the $\sigma$(Teff), we obtained the 1$\sigma$ 
uncertainty in the  slope of abundance versus excitation potential, combined with the sensitivity of the slope to T$_{\rm eff}$. 
This results in an error on the spectroscopic temperature of 60 K.

 Upon using the photometric gravities, ionization balance was not achieved for 
star 35, but the neutral and ionised iron species agreed to within 0.01 dex for star 46. This is addressed further in Sects. 4.1 and 4.3.
We reiterate, however, the uncertainty in log\,$g$, which is able to resolve the imbalance (see also Sect. 3.2).
At the metallicity of NGC 1261 of $-$1.26 dex, the GC stellar parameter study of  \citet{Mucciarelli2020} indicates that spectroscopic parameters are still reliable
and not affected by perturbations from, for example, departures from LTE. 
The final parameters for both stars are also contained in Table~1. 
\subsection{Abundance errors}
For the statistical errors on our abundances we state in 
Table~2 with our abundance results  the standard deviation and the number of  lines measured per element 
to derive its abundance ratio. 
Secondly, we estimated systematic errors in the standard manner, that is, we varied each stellar parameter about its 
 uncertainty as estimated in the previous section (T$_{\rm eff}\pm60$ K, 
log\,$g\pm0.15$ dex, [M/H]$\pm$0.1 dex,  $\xi\pm0.1$ km\,s$^{-1}$) and recorded the difference to the unperturbed parameter set. These deviations in $\log\varepsilon$ are listed in 
Table~A.1 in the appendix for both stars. 
Furthermore, an 
 identical analysis was performed, but with  the solar-scaled opacity distributions (ODFNEW), where we take one-quarter of the respective deviation  
 to reflect an uncertainty in the stars' [$\alpha$/Fe] of 0.1 dex (listed as column ``ODF'' in Table~A.1). 
 Finally, we list the total systematic uncertainty as the squared sum of all contributions is given in the last column, although 
 we caution that this is {merely a} conservative upper limit given the  
 strong correlations between  the various stellar parameters  \citep[e.g.,][]{McWilliam1995}.
\section{Abundance results}
All resulting abundance ratios and the statistical error proxies  are given in
Table~2. 
In this work,  we adopt the solar abundances of \citet{Asplund2009}.
In following Figures 2 through 5, we place these results into context with stars across 
the MW disks, bulge, and halo. References to the literature data are given in the captions.
\begin{table}[hbt]
\caption{Abundance results.}             
\centering          
\begin{tabular}{cccrcccrc}
\hline\hline       
 & [X/Fe] & $\sigma$ & $N$ & &
 [X/Fe] & $\sigma$ & $N$ & \\
 \cline{2-4}\cline{6-8}
 \rb{Species} & \multicolumn{3}{c}{Star 35} &&  \multicolumn{3}{c}{Star 46} \\
 \hline
Li\,{\sc  i} & $\llap <$$-$1.60 &  \ldots &   1  && $\llap <$$-$0.50 &  \ldots & 1$^{\rm S}  $ &  {\tiny S}  \\ 
 O\,{\sc  i} &  \phs0.71 & 0.10 &   2  &&  \phs0.54 &  0.17 & 2 & {\tiny S} \\ 
Na\,{\sc  i} &  \phs0.10 & 0.06 &   4  &&  \phs0.39 & 0.16 & 4 & {\tiny S} \\ 
Mg\,{\sc  i} &  \phs0.34 & 0.20 &   9  &&  \phs0.49 & 0.18 & 6  & \\ 
Al\,{\sc  i} &  \phs0.08 & 0.12 &   4  &&  \phs0.37 & 0.18 & 3 & {\tiny S} \\ 
Si\,{\sc  i} &  \phs0.33 & 0.20 &  18  &&  \phs0.25 & 0.23 & 14 &  \\ 
 K\,{\sc  i} &  \phs0.46 & 0.04 &   2  &&  \phs0.61 & 0.08 & 2  & \\ 
Ca\,{\sc  i} &  \phs0.33 & 0.26 &  20  && \phs0.25 & 0.22 & 19 &  \\ 
Sc\,{\sc ii} &  \phs0.00 & 0.18 &   5  &&  \phs0.08 & 0.25 & 5 &  \\ 
Ti\,{\sc  i} &  \phs0.29 & 0.27 &  40  &&  \phs0.23 & 0.24 & 25  & \\ 
Ti\,{\sc ii} &  \phs0.29 & 0.40 &  15  &&  \phs0.28 & 0.11 & 10 &  \\ 
 V\,{\sc  i} &  \phs0.17 & 0.16 &  19  &&  \phs0.01 & 0.18 & 14 &  \\ 
Cr\,{\sc  i} & $-$0.06 & 0.21 &  13  && $-$0.10 & 0.19 & 10 &  \\ 
Mn\,{\sc  i} & $-$0.35 & 0.15 &   6  && $-$0.41 & 0.12 & 6  &  \\ 
Fe\,{\sc  i} & $-$1.24 & 0.21 & 162  && $-$1.27 & 0.22 & 147  & \\ 
Fe\,{\sc ii} & $-$1.15 & 0.17 &  16  && $-$1.26 & 0.20 & 16 &  \\ 
Co\,{\sc  i} &  \phs0.03 & 0.20 &  4  &&  $-$0.07 & 0.14 & 2  &  \\ 
Ni\,{\sc  i} &  \phs0.02 & 0.27 &  36  && $-$0.04 & 0.26 & 28  & \\ 
Cu\,{\sc  i} & $-$0.41 & 0.11 &   3  && $-$0.31 & 0.15 & 2  &  \\ 
Zn\,{\sc  i} & $-$0.12 &  \ldots &   1  &&  \phs0.01 & 0.01 & 2  & \\ 
Rb\,{\sc  i} &  \phs0.11 &  \ldots &   1  &&  \phs0.20 &  \ldots & 1 &  \\ 
Sr\,{\sc ii} & $-$0.20 & \ldots & 1 && $\llap >$$-$0.70 & \ldots & 1 & \\
 Y\,{\sc ii} &  $-$0.08 & 0.34 &  5  && \phs0.08 & 0.28 & 5 & {\tiny S} \\ 
Zr\,{\sc ii} & $-$0.27 &  \ldots &   1  && \phs0.13 &  \ldots & 1& {\tiny S}  \\ 
Ba\,{\sc ii} &  \phs0.01 & 0.16 &   4  &&  \phs0.12 & 0.18 & 5 & {\tiny S} \\ 
La\,{\sc ii} &  \phs0.27 & 0.11 &   4  &&  \phs0.36 & 0.07 & 3 & {\tiny S} \\ 
Ce\,{\sc ii} &  \phs0.09 &  \ldots &   1  &&  \phs0.17 &  \ldots & 1 & {\tiny S} \\ 
Nd\,{\sc ii} &  \phs0.25 &  \ldots &   1  &&  \phs0.33 &  \ldots & 1 & {\tiny S} \\ 
Sm\,{\sc ii} &  \phs0.64 & 0.17 &   4  &&  \phs0.64 & 0.16 & 3 & {\tiny S} \\ 
Eu\,{\sc ii} &  \phs0.60 & 0.21 &   2  &&   \phs0.58 & 0.21 & 2 & {\tiny S}  \\ 
\hline                  
\end{tabular}
\tablefoot{``S'' denotes abundances that were derived from spectrum synthesis. 
Ionized species and oxygen are referenced to ionized iron.}
\end{table}
\subsection{Iron abundance}
From both stars, we find a mean metallicity of NGC 1261 of $-$1.26$\pm$0.02(stat.)$\pm$0.12(sys.), which 
is in excellent agreement with the value listed in the \citet{Harris1996} catalog. Other analyses, while differing
at the 0.1 dex level, all agree that this GC is only moderately metal poor:  
 \citet{Kravtsov2010} finds [Fe/H]=$-$1.34$\pm$0.16 and $-$1.41$\pm$0.10 dex, depending on the CMD features used, on the scale of \citet{ZinnWest1984} using their wide-field U- and 
 B-band photometry. In turn, the abundance analysis of seven stars from MIKE spectra by \citet{Filler2012} resulted in a similar value of $-$1.19$\pm$0.02 dex, which is consistent with 
the scale of \citet{KraftIvans2003}.

Ionization balance is marginally achieved for star 35 with [Fe\,{\sc i}/{\sc ii}]=$-$0.09$\pm$0.05, 
while 
both Fe species agree very well in star 46, at 
[Fe\,{\sc i}/{\sc ii}]=$-$0.01$\pm$0.05. To alleviate the discrepancy in the brighter star, a distance closer by 3 kpc or a much higher mass of $\sim$1.2 M$_{\odot}$ would need
to be imposed, neither of which are attractive options. 
While it is inconclusive from the CMD (Fig.~1), star 35 might also be on the AGB, in which case the surface gravity would decline, bringing the Fe abundances 
closer to ionization equilibrium. 
Lowering the helium mass fraction following \citet{Stromgren1982} and \citet{Lind2011NGC6397} remains unphysical.
As our error analysis shows, switching to the solar-scaled opacity distributions (ODFNEW) allows us to remove the ionization imbalance in 
star 35. However, this is not supported by the $\alpha$-enhancements in the stars that are rather compatible with the enhanced opacity distributions. 
Finally, recomputing the iron abundances with a subset of those lines with available NLTE-corrections from \citet{Bergemann2012} only led to minor
changes in [Fe/H] from either species so that departures from LTE cannot impose the balance, neither.
In the following, we continue with the {bona fide} parameter set, irrespective of the remaining ionization imbalance (see also the discussions in \citealt{Koch2008}). 
Accordingly, abundance ratios of neutral species are always referenced against neutral Fe and those from ionized species against Fe\,{\sc ii}. 
Likewise, O\,{\sc i} is referenced against Fe\,{\sc ii} since the two species are the 
dominant form and  have similar gravity sensitivity.  
\subsection{Light elements: Li, C, O, Na, Al, K}
Both lithium and carbon are barely
detected in both stars; the limits we place on their abundance
ratios are typical for luminous red giants 
in the field and in GCs \citep{Lind2009,Placco2014,Kirby2015}. {We note, however, the presence of a Li-rich giant in this GC \citep{Sanna2020}}. 

As for oxygen, we note that  its abundance was derived with explicit care of the underlying Ca-autoionization feature of the 6363\AA\  line and 
potential weak telluric absorption affecting either used line. 
The light elements O and Na are of great interest since their abundance ratio can be
strongly affected by proton-burning reactions, a chemical signature of
multiple GC subpopulations \citep[e.g.,][]{Carretta2009NaO}.
As shown in the  left panel of Fig.~2 these elements also vary significantly in NGC~1261. 
While strongly enhanced in oxygen, the approximately solar value of [Na/Fe] in star 46
can still be compatible with a second-generation star. Similarly, star 35 has both elevated Na and O levels.
\begin{figure}[htb]
\centering
\includegraphics[width=1\hsize]{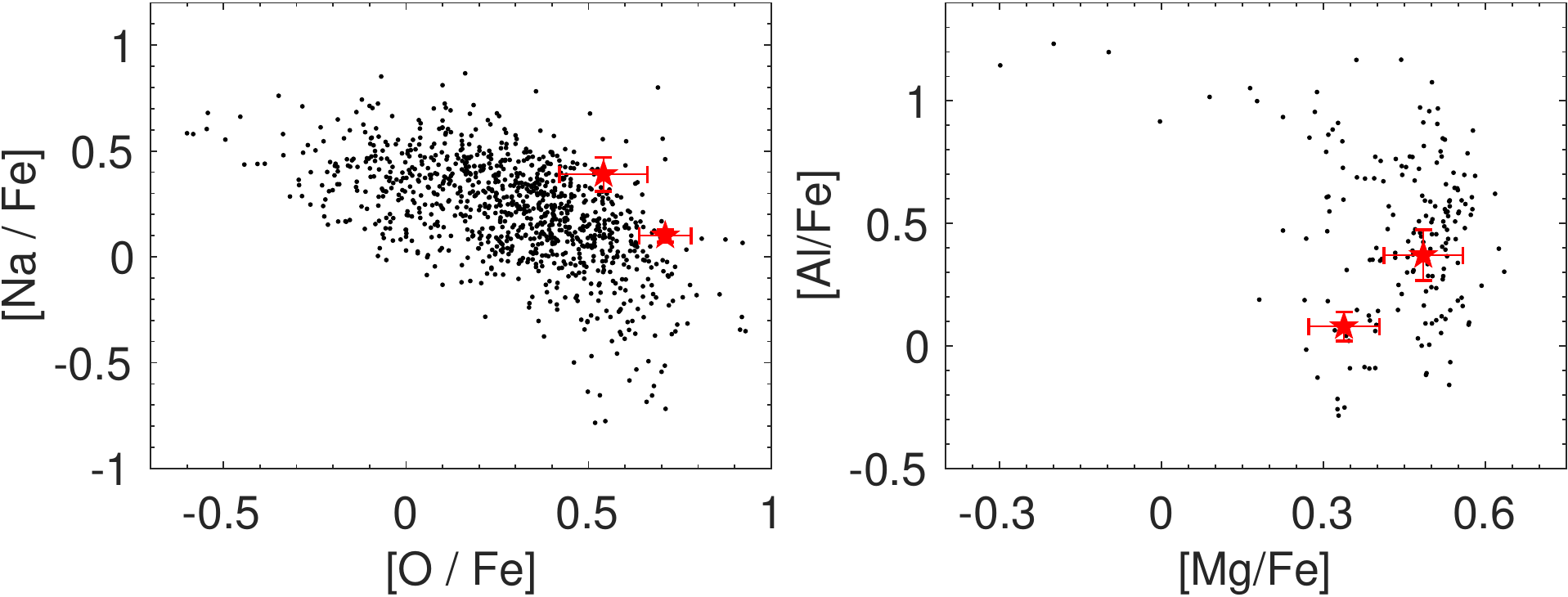}
\caption{Light element variations in NGC~1261 (red symbols) and the compilation of stars in 17 GCs  from  \citet{Carretta2009NaO}.}
\end{figure}

Mg and Al abundances show large, correlated, dispersions in GC stars
\citep[e.g.,][]{Carretta2012}, which is consistent with our results for
NGC 1261 (right panel of Fig.~2). Unlike O and Na, however, the Al and Mg
abundances are positively correlated.  We simply conclude that the presence
of abundance variations in NGC~1261, measured in this work, are in line with the finding
of \citet{Filler2012} and we attribute the abundance differences to different GC subpopulations. 
Given the small size of our sample, we refrain from uniquely associating our stars with either first or second generation. 
\subsection{$\alpha$-elements: Mg, Si, Ca, Ti}
\begin{figure}[htb]
\centering
\includegraphics[width=0.9\hsize]{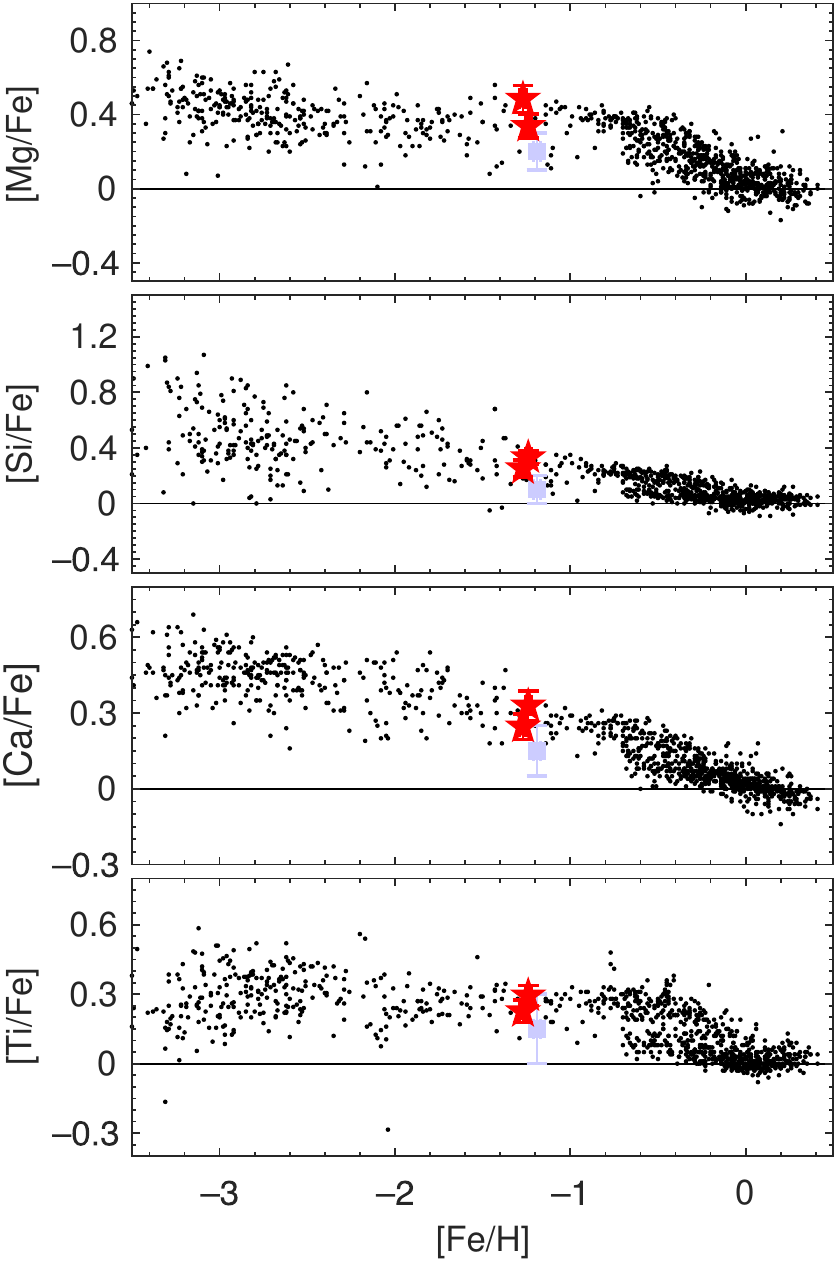}
\caption{Abundance results for the $\alpha$-elements. Milky Way halo \citep{Roederer2014}
and disk \citep{Bensby2014} stars are shown as black dots,{ and the mean values for NGC 1261 by \citet{Filler2012} are indicated
in light blue.}}
\end{figure}

The abundance ratios of the $\alpha$-elements are shown in Fig.~3. We note that ionization equilibrium in Ti follows that of iron. 
{While 
$\log\varepsilon$\,(Ti\,{\sc i})$-\log\varepsilon$\,(Ti\,{\sc ii})  
takes values of $-$0.09 and $-0.06$ in stars 35 and 46, respectively, 
the abundance ratios relative to the respective species (viz. [Ti\,{\sc i}/Fe\,{\sc i}] and  [Ti\,{\sc ii}/Fe\,{\sc ii}]) are consistent to within the errors (see Table~2).}

If we simplistically form a straight average of the four elements we measured, we find a mean [$\alpha$/Fe] of 0.32 (star 35) and 0.30 dex (star 46).
Overall, all elements show very low star-to-star scatter save for Mg, which we assign to the occurrence of multiple populations in this GC. 
These elevated values are in contrast to the numbers reported by \citet{Filler2012}, who measured $\alpha$/Fe ratios that are significantly lower than those
found in halo stars of similar metallicity. Their study suggested values for [Mg,Si,Ca,Ti/Fe] of $\sim$0.1 dex contrasting with the canonical halo plateau 
value of $\sim$0.4 dex. Combined with the younger age of the cluster, the $\alpha$-depletion with respect to systems at similar metallicities claimed by \citet{Filler2012} 
could suggest an accretion origin from a dwarf galaxy because those tend to have systematically lower [$\alpha$/Fe] ratios, owing to their low star formation 
efficiencies \citep[e.g.,][]{Matteucci1990,Koch2008Carina,Tolstoy2009}. We, however, revisit a possible accretion origin of NGC~1261 
based on its actual $\alpha$- and heavy element composition in Sects.~5 and 6.
\subsection{Fe-peak elements: Sc, V, Cr, Mn, Co, Ni, Cu, Zn}
Abundances for the eight (nine including Fe) iron-peak elements we were able to measure are shown in Figs.~4 and 5
\begin{figure}[htb]
\centering
\includegraphics[width=0.9\hsize]{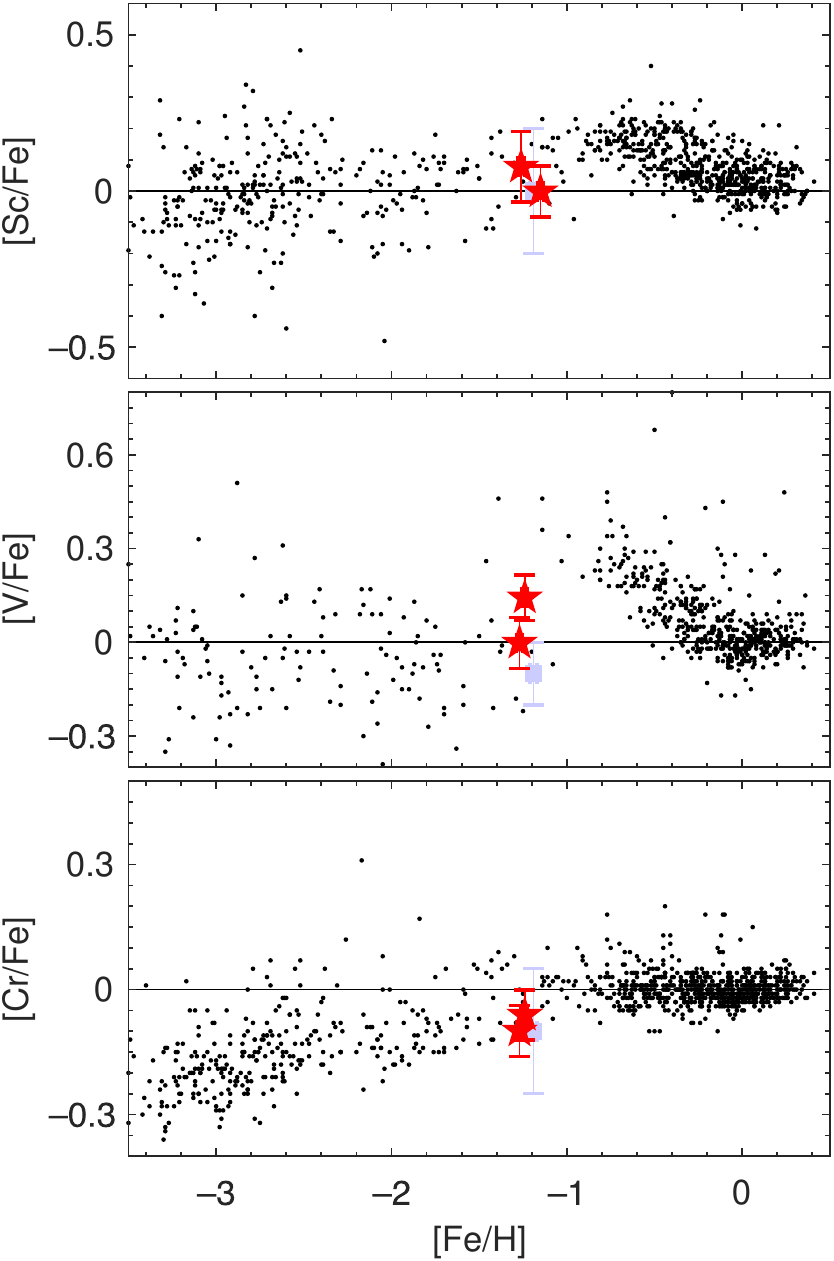}
\caption{Same as Fig.~3, but for Fe-peak elements. The  values for Sc and V in the disk are   from \citet{Battistini2015}.}
\end{figure}
\begin{figure}[htb]
\centering
\includegraphics[width=0.9\hsize]{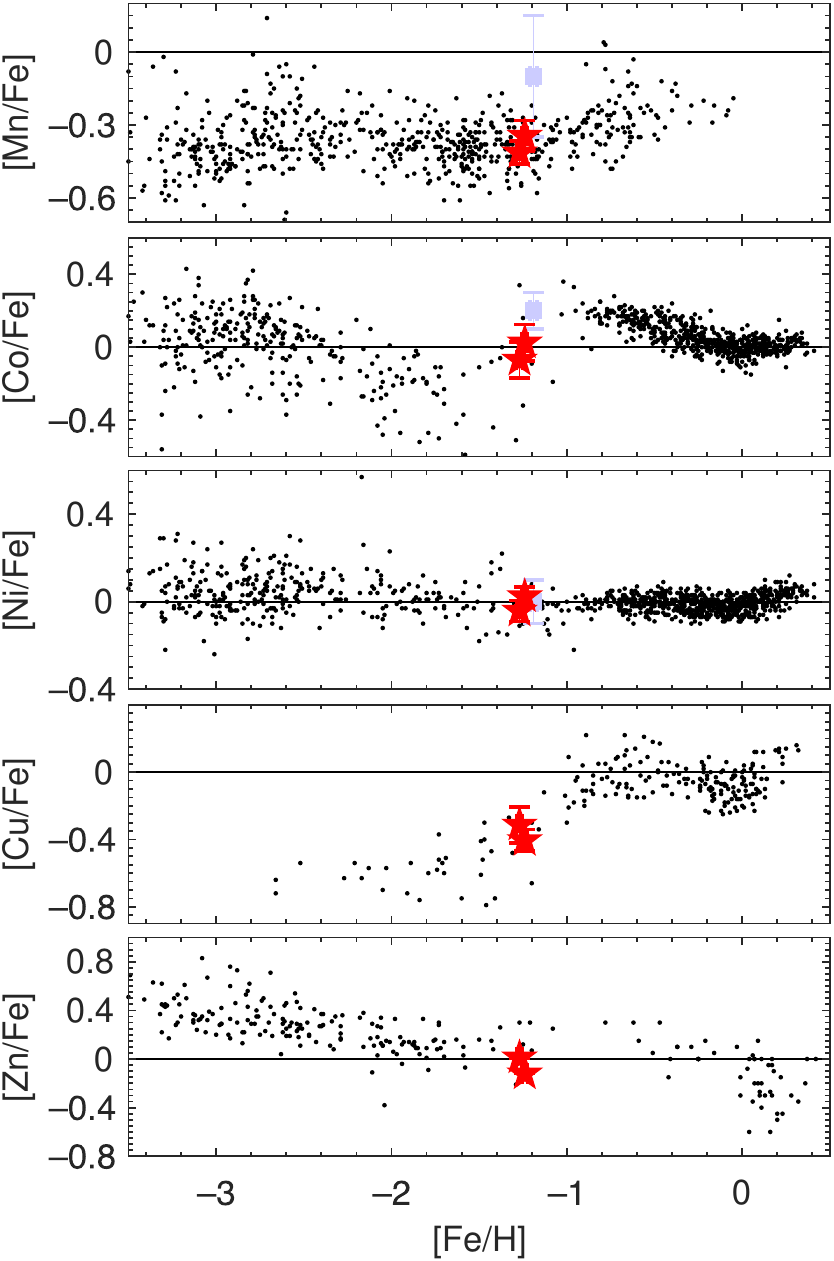}
\caption{Same as Fig.~3, but for the remaining Fe-peak elements. The data for Mn were taken  from \citet{Sobeck2006}, Co abundances from  \citet{Battistini2015}, and 
Cu in the MW halo and disks from the studies of \citet{Mishenina2002,Mishenina2011}. The reference sample of Zn abundances in the MW is  from \citet{Barbuy2015}.}
\end{figure}
There is little to notice in the distibution of these elements and suffice it to say that they all follow the trends outlined by moderately metal-poor halo field stars, 
manifesting the origin of the Fe-peak in Supernovae (SNe) of type Ia.
Both stars have similar abundance ratios and
most values are also broadly consistent with the 
study of \citet{Filler2012}.
\section{$r$-process dominance}
We were able to determine abundance ratios of the neutron-capture elements  Rb, Sr, Y, Zr, Ba, La, Ce, Nd, Sm, and Eu. 
The values for Sr are rather to be taken as lower limits because of the strong saturation of the used 4215 \AA\  line 
and the strong blending and uncertainty in the continuum placement owing to the low S/N ratio. We note, however, that the results from EW measurements 
and spectral synthesis agree.
Ba abundances were determined from the 4554, 5853, 6141, 6496 \AA~lines, while 
our [Eu/Fe] is based on the red lines at 
6437 and 6645 \AA. 

A subset of the heavy element abundances are shown in Fig.~6. 
\begin{figure}[htb]
\centering
\includegraphics[width=0.9\hsize]{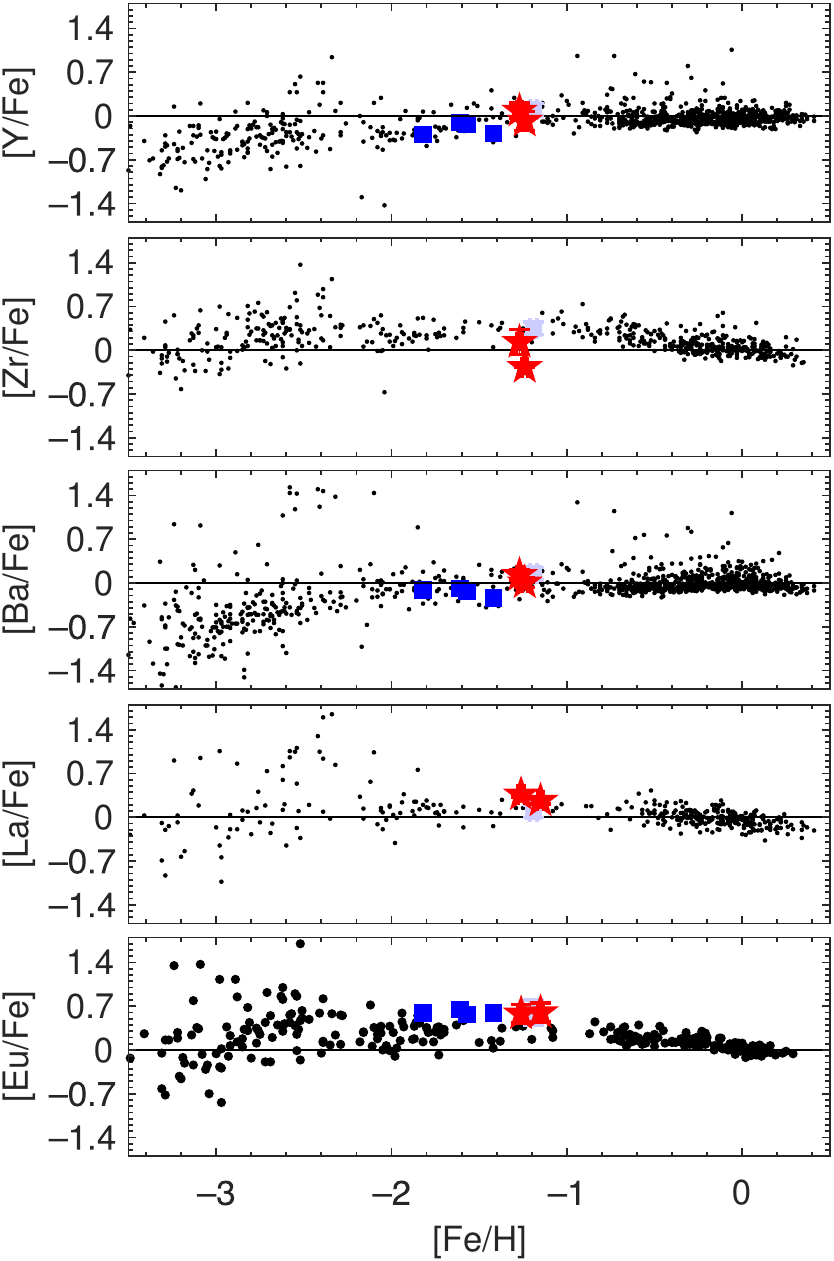}
\caption{Same as Fig.~3, but for the neutron-capture elements ($Z>$30). 
Zr and La abundances of disk stars are from \citet{Battistini2016}. 
Here, all axes are on the same scale so as to facilitate comparing the enhancements.\ The
$r$-rich stars associated with Gaia Enceladus are shown as blue squares \citep{Aguado2021}.
}
\end{figure}
While Zn, Y, and Ba are consistent with the approximately solar values as seen in the halo field, we note a slight enhancement in [La/Fe] at the 0.3-dex level.
Most strikingly, however, the [Eu/Fe]  ratio is very high -- at 0.6 dex. This is bolstered by the { work} of \citet{Filler2012}, who report an even 
larger overabundance 0.7 dex from their seven stars. We note that our values have not been corrected for NLTE.  \citet{Mashonkina2014} 
estimate Eu-corrections on the order of 0.1 dex, 
albeit for a more metal-poor sample, which would result in even larger [Eu/Fe] ratios in our stars.

In order to further investigate the origin of the heavy elements in NGC~1261, we juxtapose in Fig.~7 the solar-scaled $s$- and $r$-process 
contributions \citep{Burris2000}. The overlap of the cluster stars and the $r$-process track is striking, and the full heavy-element abundance pattern 
can best be reproduced (in a least-squares sense) by an almost pure $r$-pattern with a 5--10\% admixture of $s$-process yields. 
\begin{figure}[htb]
\centering
\includegraphics[width=1\hsize]{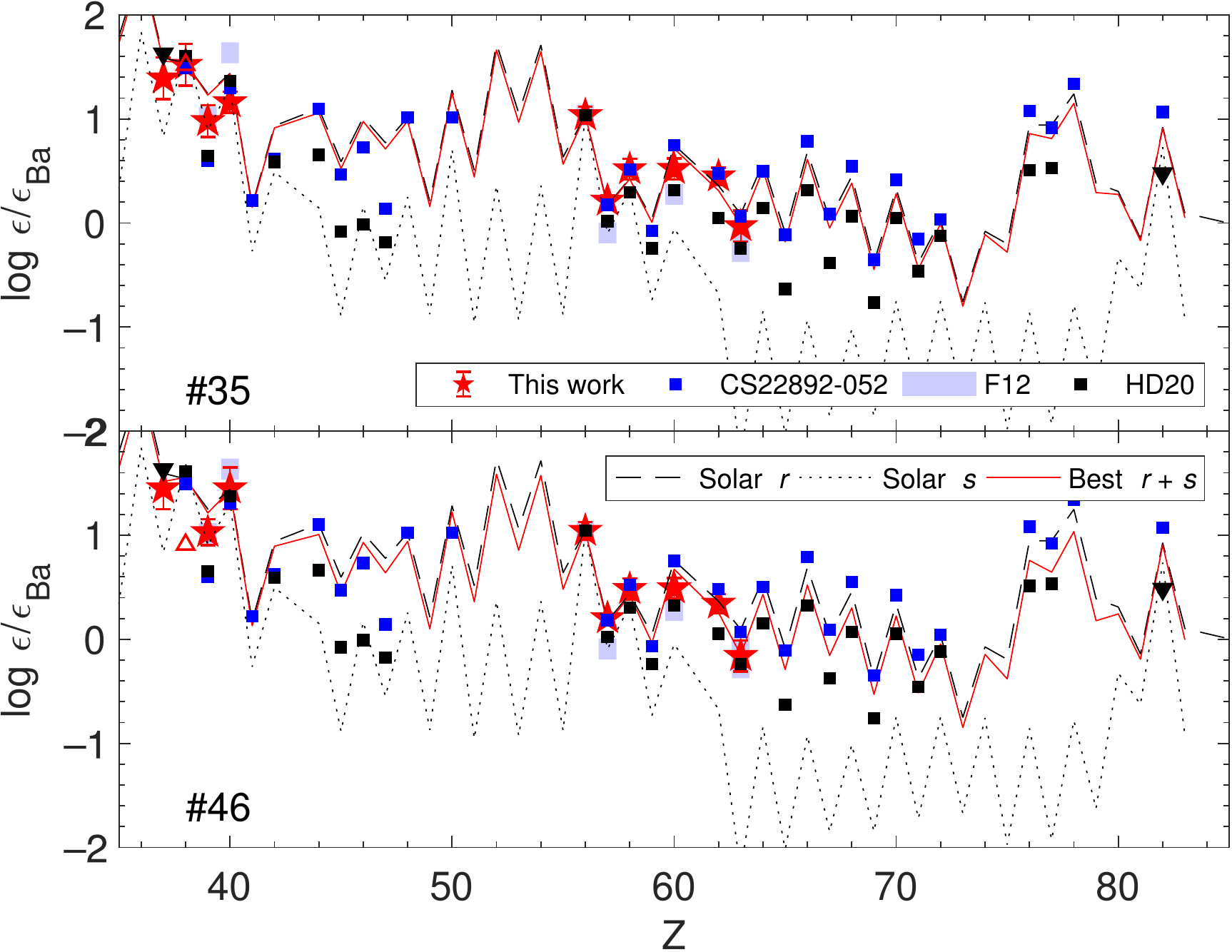}
\caption{Heavy element distributions of both target stars. Also shown are the solar-scaled curves from \citet{Burris2000} and the $r$-rich benchmark stars CSS22892$-$052 and HD20 
\citep{Sneden2003,Hanke2020HD20}. The mean and standard deviations of the NGC~1261 sample of \citet[][``F12'']{Filler2012} are indicated as blue shaded boxes. 
All patterns are normalized to Barium.}
\end{figure}
The $r$-process dominance of NGC~1261 is further highlighted when comparing its patterns to metal-poor halo field star CS22892$-$052 \citep{Sneden2003},
which is governed by $r$-process nucleosynthesis.

The  [Ba/Fe] and [Eu/Fe] ratios of the stars qualify them as 
r-I stars\footnote{Such { r-I} stars are classified via 0.3$\le$[Eu/Fe]$\le$1.0 and [Ba/Eu]$<$0.
{ Similarly, the strongly enhanced r-II stars are defined via  [Eu/Fe]$>$1.0 and [Ba/Eu]$<$0 \citep{BeersChristlieb2005}.}
}, although those are  
primarily found at lower metallicities \citep{Westin2000,BeersChristlieb2005}\footnote{We also note the existence of r-I stars at moderately low metallicity in the Galactic bulge \citep{Johnson2013}.}. 
Comparing our NGC 1261 stars to two well-known r-II (CS 22895-052) and r-I (HD20; \citealt{Hanke2020HD20}) stars, a clear $r$-process { trend lying between} these two r-I and r-II stars is seen despite the higher metallicity of stars 35 and 46. Surprisingly, even the lighter (37$\le$Z$<$50) elements seem to follow a smooth $r$-process trends.

\subsection{Lighter $n$-capture elements (Rb, Sr, Y,  Zr)}
The lighter $n$-capture elements (Rb, Sr, Y, and Zr) can be produced in weak $s$- and $r$-processes as well as their main channels. 
The Rb abundance can be used as a 
tracer of the $n$-density or $n$-exposure in the n-capture environments \citep[e.g.,][]{Pignatari2010}. Classically speaking, Sr, Y, and Zr are predominantly formed in the $s$-process in 
the solar inventory. However, the large star-to-star scatter detected in these elements \citep[e.g.,][]{Francois2007,CJHansen2012} 
 indicate that several processes may contribute to their production at the metal-poor end of the metallicity scale (see also \citealt{McWilliam1998}). 
 Studies have shown that a lighter element primary process \citep[LEPP; see, e.g.,][]{Busso1999,Travaglio2004,Bisterzo2012,CJHansen2014}, including the alpha-rich freeze-out \citep{Woosley1992}, can contribute to their production, while 
 other yield computations may be able to explain a larger contribution through model effects such as rotation in AGB stars \citep[e.g.,][]{Piersanti2013,Cristallo2015}. 
 The weak $s$-process is typically associated with fast-rotating massive stars \citep[FRMS;][]{Hirschi2007,Frischknecht2016, Choplin2018}. The fast rotation in combination with C-
 reactions leads to the $n$-excess that  facilitates a weak $s$-process, which, relatively speaking, produces more Sr, Y, Zr than Ba and La. 
 Hence, the ratio of Y/La may be used to trace weak versus main $s$-process contributions, but it may also be attributed to the mass and metallicity of the AGB donor 
 \citep{Busso1999}. Recently, Sr was 
 shown in the direct observations of kilonova AT2017gfo to be produced in the merger of two neutron stars \citep{Watson2019}. 
 Hence, as most $n$-capture processes need to pass the A$\sim90$ region, a complex mixture of contributions from various processes can be expected in the formation of Rb-Zr. 
Finally, the Galactic GC M15 ([Fe/H]$\sim -2.4$) is included in our comparison that follows owing to its heavy ($r$-process) element content, where traces of a scatter in the light 
$n$-capture elements, and even more so in the heavy elements, have been found  \citep{Sneden1997,Otsuki2006,Worley2013}. This indicates contributions from several processes forming light and heavy elements in differing amounts, or it calls for incomplete mixing.
\subsection{Source of the Eu overabundance}
Despite the variance
between the light element abundances, the similarity of the [Eu/Fe] ratios in our two stars is consistent with the possibility that 
the $r$-process enhancement in NGC~1261 is primordial.  Clearly, chemical analysis 
of many more than two stars in the cluster is required to test this idea.

If the $r$-process enhancement is of primordial origin, then
the excess mass of Eu in the cluster can be estimated from the enhancement
of the NGC~1261 [Eu/Fe] ratio, at $+$0.6 dex, which is approximately 0.3 dex higher
than typical MW stars with similar [Fe/H], combined with the expected
total initial mass of stars in NGC~1261, at $1.32\times10^{6} M_{\odot}$ 
\citep{Webb2015}.  In this way, we find an excess europium mass of
$2\times10^{-5}$ M$_{\odot}$, assuming that the total star formation efficiency 
of the primordial molecular cloud was close to unity. 
 This europium mass
is close to the predicted ejected r-process masses, at $3\times10^{-5}$ and
$5\times10^{-5}$ M$_{\odot}$, for two individual neutron star-neutron star (NS) merger cases studied by
\citet{Goriely2011}.  If we adopt the europium $r$-process mass fraction of
0.01, implicit in \citet{Goriely2011}, we estimate a total $r$-process mass
for NGC~1261 of $2\times10^{-3}$ M$_{\odot}$, which is similar to a predicted
total mass of $r$-process ejecta for NS merger scenarios in the
range $10^{-3}$ to $10^{-2}$ M$_{\odot}$
(e.g., \citealt{Goriely2011,Shibata2019}, and references therein).

If the total star formation efficiency of the NGC~1261 parent molecular cloud
was not close to unity, then the measured r-process abundances might be explained
by multiple NS merger events or an upward revision of the predicted r-process yields 
{ \citep[e.g.,][]{Molero2021}}.
We note that the predicted mass of ejected r-process material for NS mergers is
sensitive to the equation of state of the NS, the mass ratio of the binary, and
whether the merger is with another NS or a black hole  \citep[e.g.,][]{Shibata2019}.

We finally note that magneto-rotational SNe could also be a possible site of the $r$-process material, in which case the higher $\alpha$-abundance could be easily explained. 
However, as seen in \citet{Reichert2021,Reichert2021MRSN} there might not be a direct link between the $\alpha$- and $r$-process enhancement 
and in such SNe the $r$-process Eu production might be a bit lower ($\sim5\times10^{-6}$ M$_{\odot}$).
\subsection{Comparison with $s$- and $r$-rich globular clusters}
\subsubsection{Origin of the heavy elements}
\begin{figure*}[htb]
\centering
\includegraphics[width=0.52\hsize]{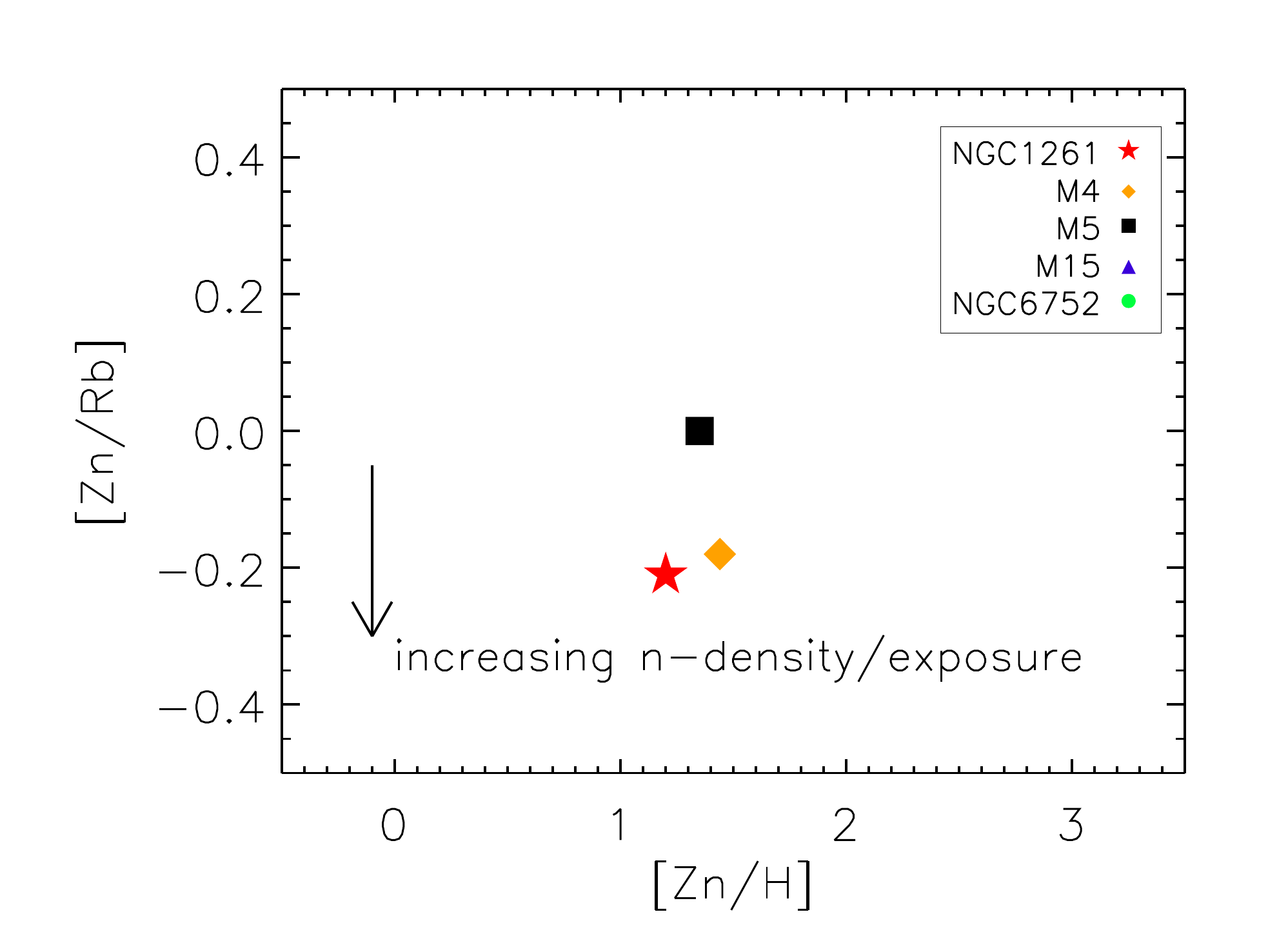}\hspace{-1cm}
\includegraphics[width=0.52\hsize]{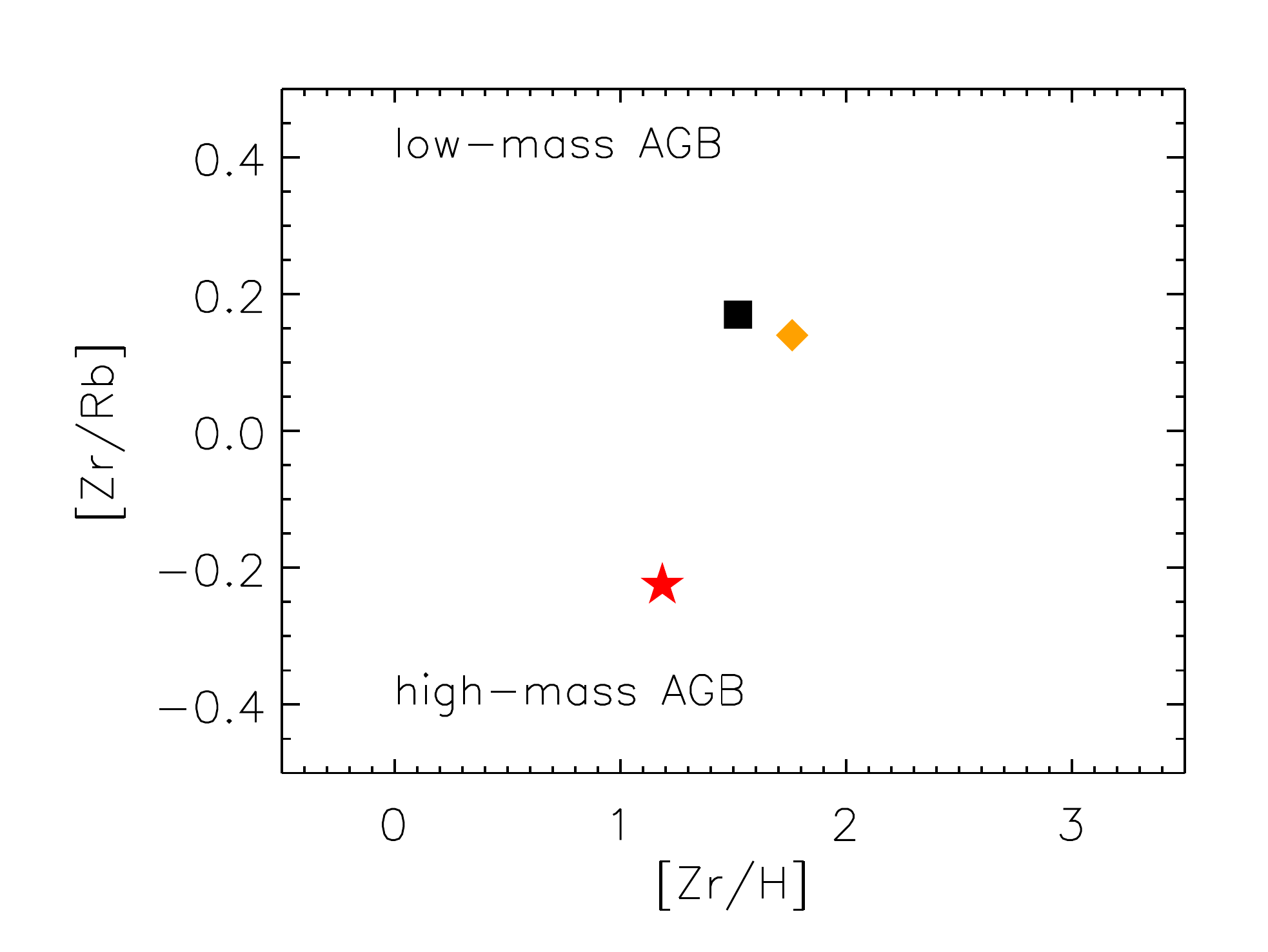}\vspace{-0.5cm}
\includegraphics[width=0.52\hsize]{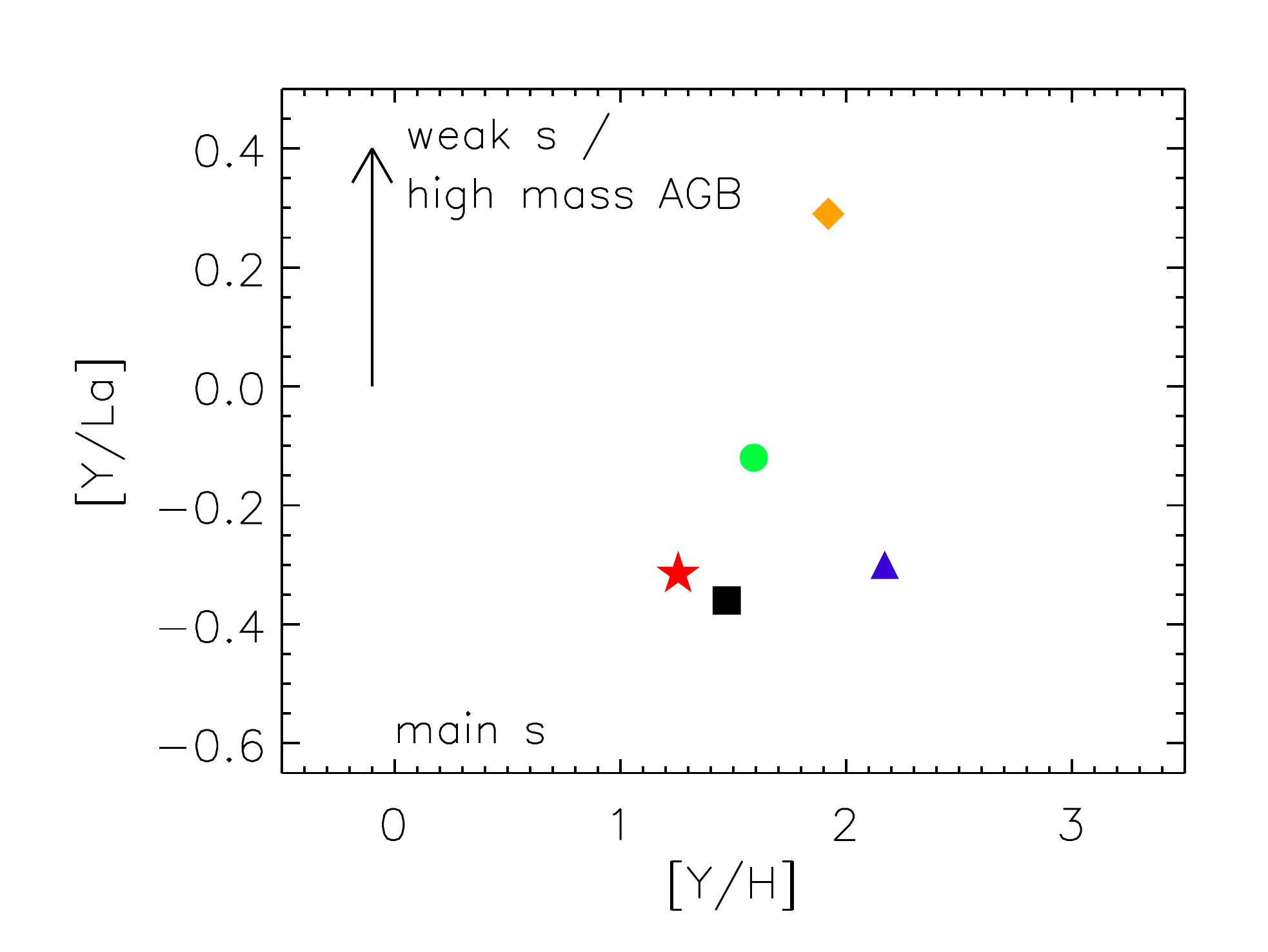}\hspace{-1cm}
\includegraphics[width=0.52\hsize]{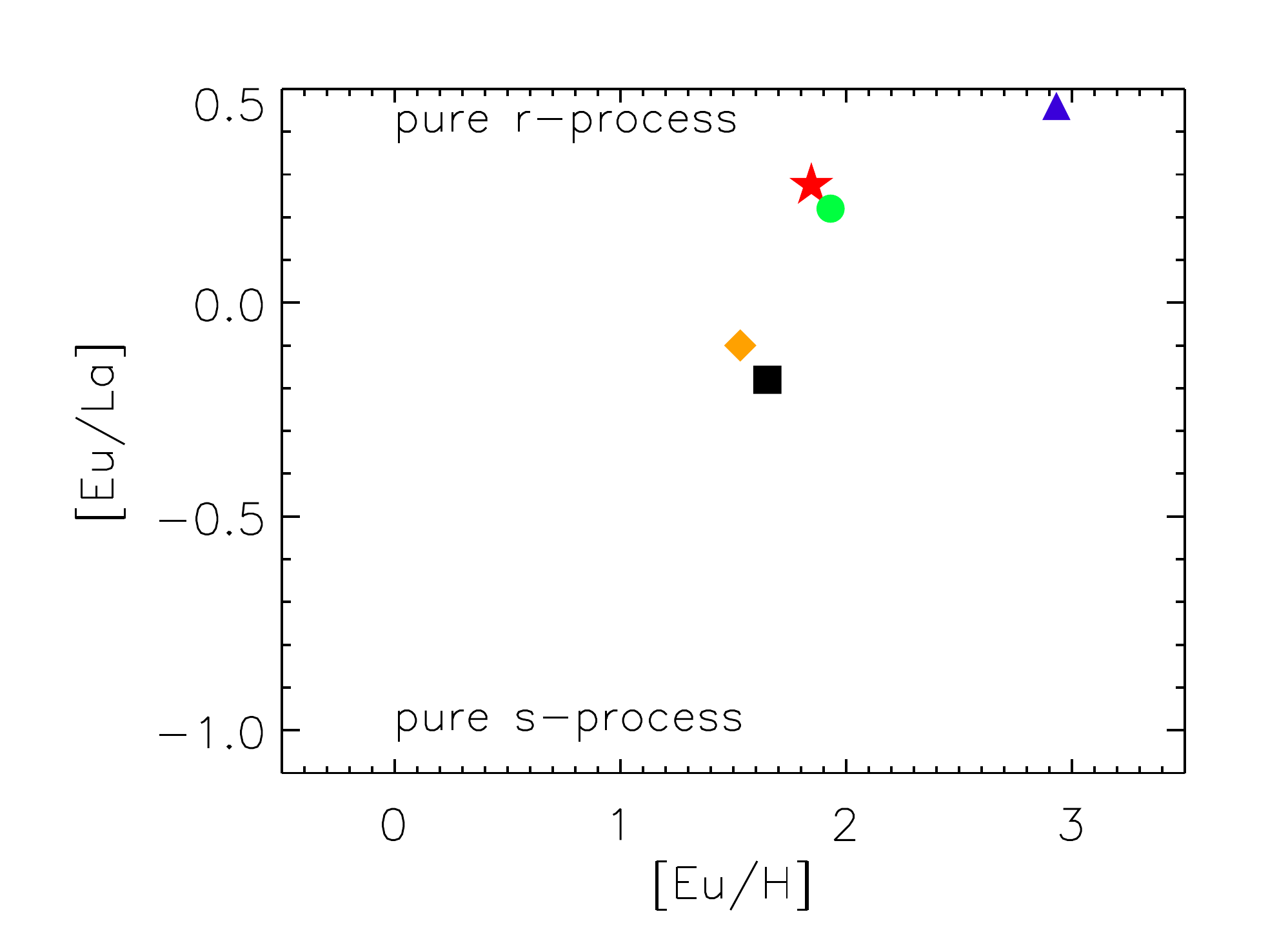}
\caption{Abundance ratios for selected elements that are tracers of the origins of the $n$-capture elements in 
NGC 1261 and reference GCs. Shown are the average values of the respective stellar samples{.
 The illustrative trends and nucleosynthetic distinctions in each panel are based on computations from 
\citet[][Zn/Rb]{Pignatari2010}, 
\citet[][Zr/Rb]{Cristallo2015fruity}, 
\citet[][Y/La]{Frischknecht2012}, and
\citet[][Eu/La]{Arlandini1999}. 
See main text for discussions.}
}
\label{fig:heavy}
\end{figure*}
Figure~\ref{fig:heavy} shows selected heavy element ratios (Zn/Rb, Zr/Rb, Y/La, and Eu/La)\footnote{ We note that we explicitly chose to 
avoid Fe in any of these figures because it clouds the $r$- and $s$-process traces we seek \citep[see, e.g.,][]{CJHansen2012}.} 
from which we can gain insight into the heavy element enrichment of stars 35 and 46. 
For comparison, we selected clusters of similar metallicity with well-populated chemical abundance patterns and with clear $r$- and $s$-process dominance. In this case, M4 and M5 
\citep{Yong2008a,Yong2008b} have a matching [Fe/H] of $\sim$$-$1.2 dex and a stronger $s$- and $r$-process contribution, respectively. 
Another GC with a scaled $r$-process pattern is the slightly more metal-poor NGC~6752 ([Fe/H]$\sim$$-1.6$), where the elements between Ba and Dy are seen to follow a solar-scaled, $r$-process distribution \citep{Yong2005}.

The [Ba/Eu] ratio is oft-used to provide information on the relative $s$/$r$ ratio in the stars \citep[e.g.,][]{Spite1992,CJHansen2018}. The high values of  [Eu/Ba] and [Eu/La]  strongly 
indicate that the stars in NGC~1261 are $r$-process dominated (see Fig.~\ref{fig:heavy}), {where the ``pure'' limits are placed 
according to $s$- ([Eu/La]$_s\sim-1.0$) and $r$-fractions ([Eu/La]$_r\sim0.4$), computed following \citet{Arlandini1999}.}
In turn, 
the observed low [Zn/Rb] and [Zr/Rb] ratios could indicate a high neutron density or exposure leading to a slight relative overabundance of Rb (at [Rb/Fe]$\sim0.15$ dex). { To this effect, we also illustrate the trends of varying respective model parameters in Fig.~8.
The weak-$s$ computations in \citet{Pignatari2010} indicate that increasing $n$-exposure and -density ultimately increase the Rb and other $s$-process element abundances. In particular, using solar abundances and the production factors from their Table~9 shows that [Zn/Rb] reaches from just subsolar to $\sim -0.3$ dex with increasing $n$-density and -exposure. The exact value depends on the model employed.}

In particular,  [Zr/Rb] suggests a  contribution from more massive AGB stars \citep{Yong2006,Yong2008a} 
in NGC 1261, as we also verified by comparison with the AGB yields of the F.R.U.I.T.Y. database \citep{Cristallo2011,Cristallo2015fruity}. { A 1.3\,M$_{\odot}$ AGB model with $Z$=0.001 and no rotation yields a [Zr/Rb] ratio of $\sim 0.4$ dex, while a more massive (5\,M$_{\odot}$) AGB 
with the same $Z$ yields a much lower value ($\sim -0.4$ dex).}
The high [Y/La] seen in M4 clearly confirms its stronger $s$-process contribution, while the low ratios in NGC 1261, M5, and M15 indicate 
a relatively low $s$-process contribution, which may originate from the weak s or, alternatively, from metal-poor  AGB stars of higher mass. 
A strong main $s$-process contribution would be expected for a 2\,M$_{\odot}$ AGB star (with the same Z as given above), leading to [Y/La]$\sim-0.6$. 
Overall, the AGB mass is generally the 
dominant $s$-process nucleosynthesis govenor, 
while  metallicity is normally a mere secondary parameter as shown in, for example, \citet[][their Figs.~12, 13]{CJHansen2018}.

Typical light $s$-to- heavy $s$ ratios in FRMS would, via the weak $s$-process, 
produce values greater than zero as is predicted for [Sr/Ba]$>0$ \citep[see, e.g.,][]{Frischknecht2012}.
This is in excellent agreement with the high [Eu/La] detected in NGC 1261, which is only rivaled by the more metal-poor M15. 
A slight $s$-process contribution is expected at  the moderately high [Fe/H]$\sim-1.2$ of NGC 1261, and it is therefore not surprising that M15 would show a 
cleaner and stronger $r$-process imprint.
Finally, we note that the intermediate neutron capture process ($i$-process; \citealt{CowanRose1977}) is unlikely as the source of the 
heavy elements in NGC 1261 because that would lead to higher positive [Ba/La] ratios  \citep{Hampel2016,Reichert2021}, while our measured value is negative.
\subsubsection{Heavy-element spreads}
\begin{figure}[htb]
\centering
\includegraphics[width=1\hsize]{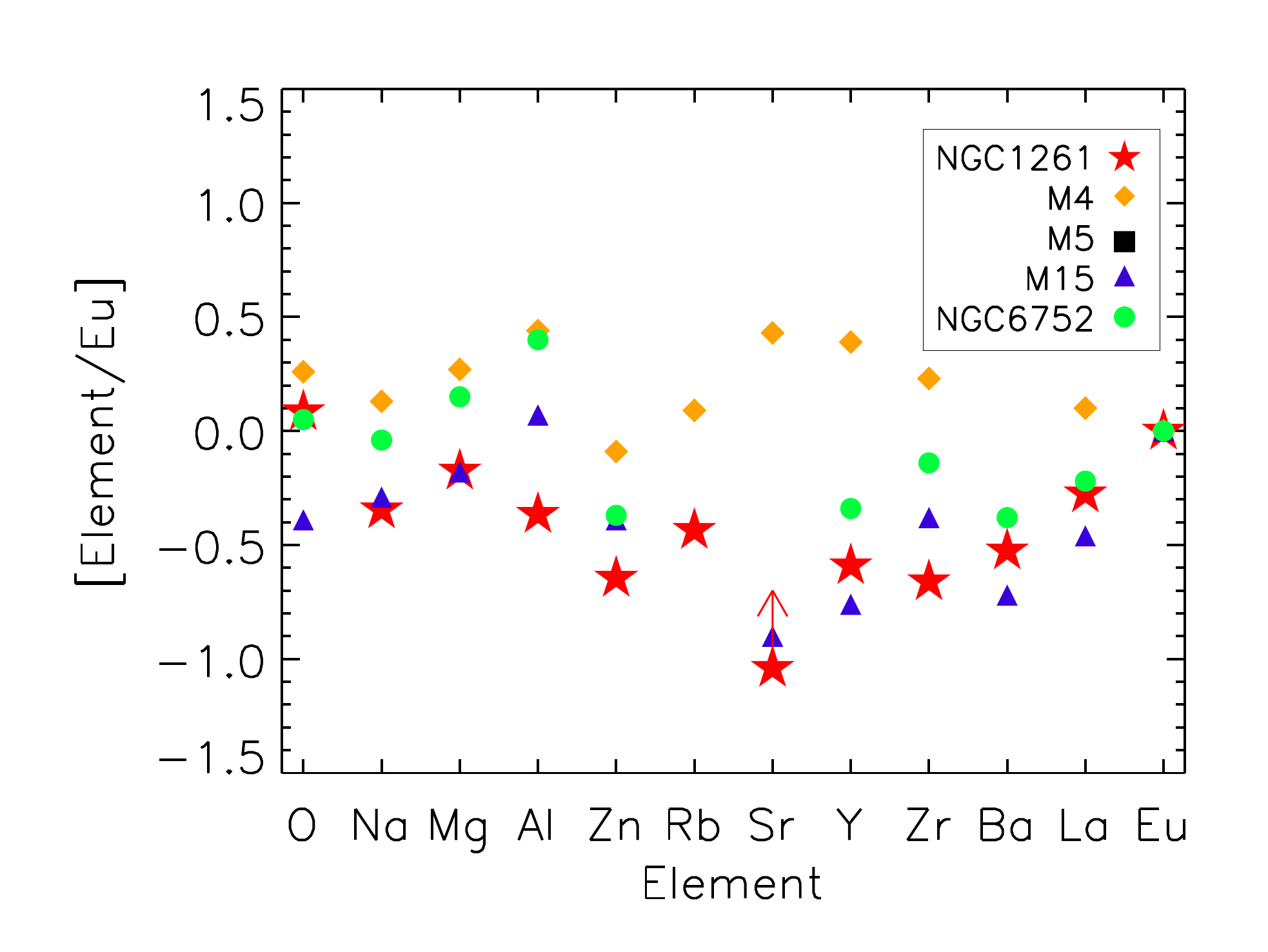}\vspace{-0.5cm}
\includegraphics[width=1\hsize]{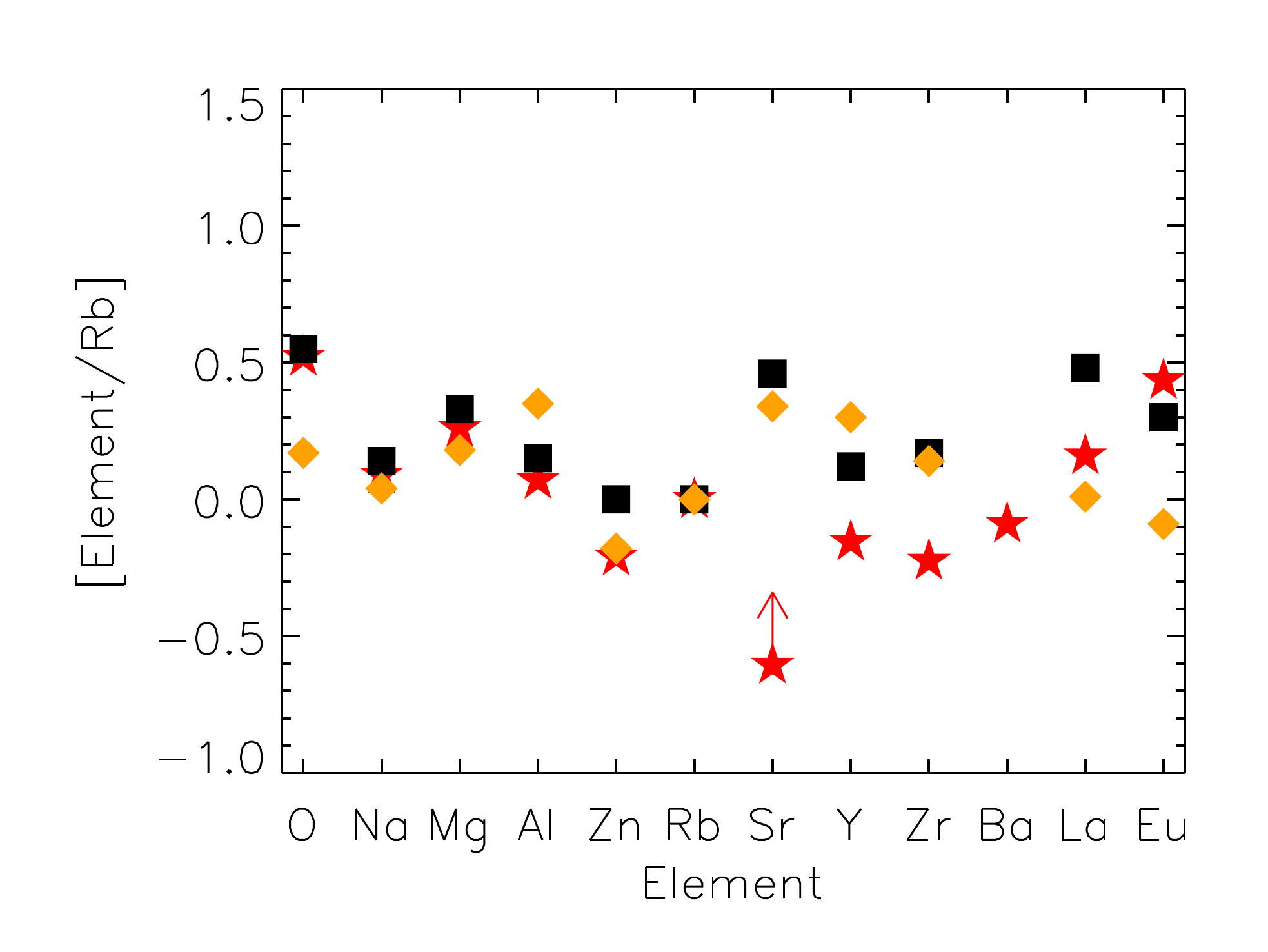}\vspace{-0.5cm}
\includegraphics[width=1\hsize]{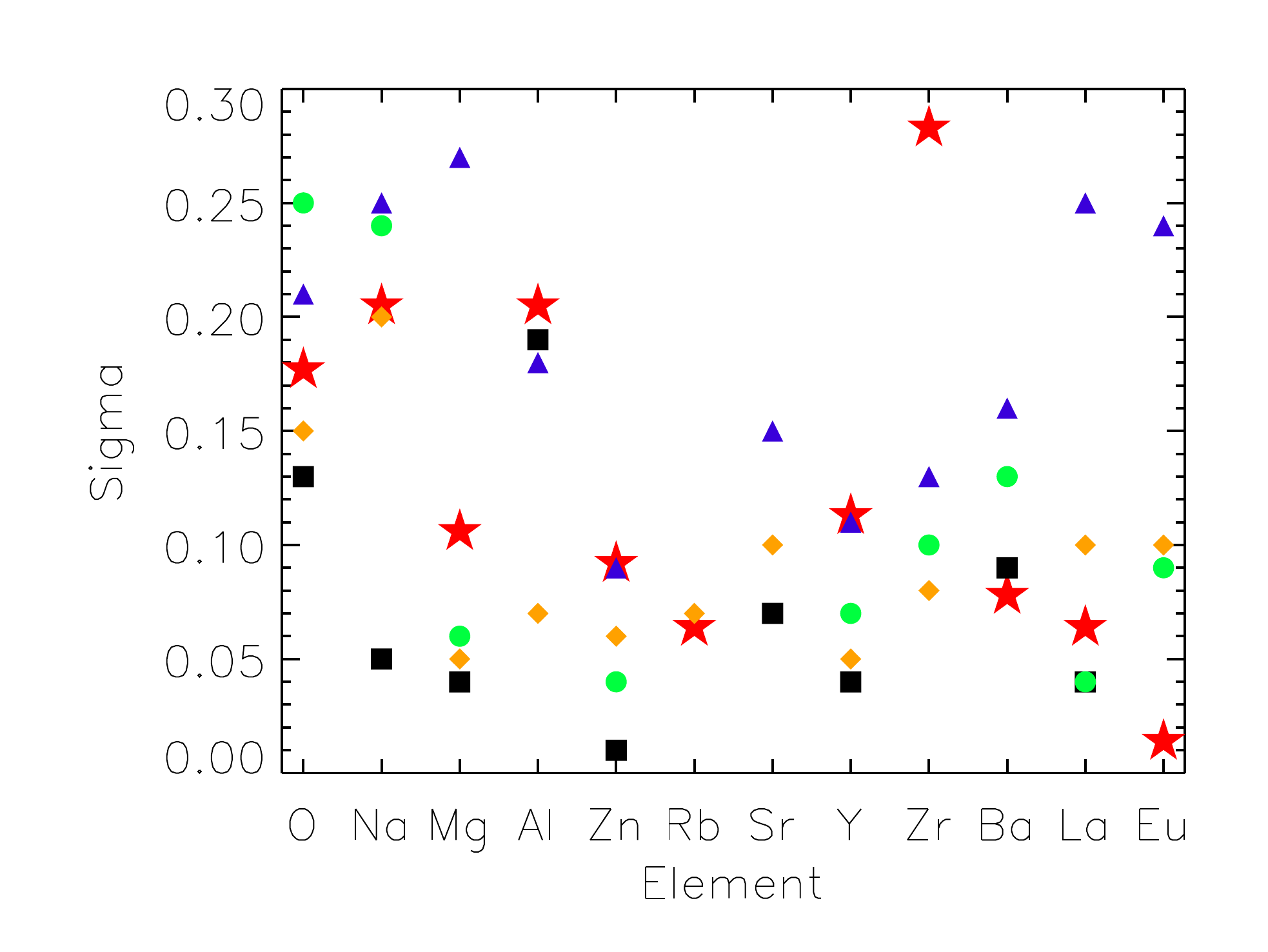}
\caption{Selected element ratios relative to Eu (top panel) and Rb (middle) for NGC 1261 and the reference GCs. The bottom panel indicates the 
1$\sigma$ star-to-scatter in each GC.}
\label{fig:scatter}
\end{figure}
Various elements and their line-to-line standard deviation are shown in Fig.~\ref{fig:scatter}. The top panel shows element abundances normalized to Eu (r-process), the middle panel 
shows abundances scaled to Rb (weak s), and the bottom panel shows the 1$\sigma$ star-to-star scatter. 
When scaling to the $r$-process element Eu, a large scatter is seen between the lighter-to-heavy element ratios as is also common in metal-poor field stars 
\citep{Roederer2018,CJHansen2012}. However, strong differences also occur between the more $s$-rich M4 and the other $r$-process dominated clusters. 
The largest difference with respect to Eu is seen for Sr, where Sr in NGC 1261 is only a lower limit, thereby decreasing this spread likely to the level of the Y/Eu variation 
if a detection on Sr could be placed in NGC~1261. 

The [element/Rb] ratio tracing a lighter element (likely in a weak $s$-process) 
appears to show less of a spread amongst the GCs. We note, however, that fewer clusters have Rb detections, so this comparison may not { be fair} or complete. 
The lighter elements O and Na are known for their intra-cluster spreads, which is related to the occurrence of multiple populations in the GCs. However, the 
bottom panel of Fig.~\ref{fig:scatter} indicates that larger scatters are also seen in some heavier elements.
 For several elements and in several clusters, a large variation is detected, notably for Zr in NGC 1261, but also in the heavy elements like Eu. 
 The latter  is mainly driven by M15, which is known to have a high degree of heavy $n$-capture spread \citep{Sneden1997,Otsuki2006,Worley2013}.
Finally, a high Zr abundance  and spread could also be produced if a weak $r$-process were also active in NGC 1261, as discussed in \citet{CJHansen2012},  and may as such not point directly toward the mass of the AGB.
 In particular,  the numerous processes that contribute to the region around  A$\sim$90 complicate tracing their production at such relatively high metallicities.
\section{Discussion and conclusions: The origin of NGC~1261}
Based on ages and orbital properties, \citet{Massari2019} associated NGC~1261 with the Gaia-Enceladus merger event \citep{Helmi2018GaiaEnceladus,Belokurov2018},
which not only built a major fraction of the Galactic inner and outer halos \citep[e.g.,][]{Naidu2021} but must also have lost parts of its GC system to 
the outer halo \citep[e.g.,][]{Kruijssen2019}.  
An accretion origin of this GC is  in line with its younger age \citep{Marin-Franch2009,Kravtsov2010}. 
In particular, \citet{Massari2019} asserted that all but one of the 28 GCs dynamically associated with Gaia-Enceladus have apocenters closer than 25 kpc, which also
holds true for NGC~1261, where
\citet{Baumgardt2019} found a close pericenter of 1.4 kpc and an apocenter distance of 20 kpc. 

Thus far, the merger-GC ties are likely, but to further sculpt this picture, the addition of a chemical abundance space is imperative \citep[e.g.,][]{Nissen1997,NissenSchuster2010}.
In Fig.~10, we overplot the mean $<$[$\alpha$/Fe]$>$ ratio, defined as the straight average of the Mg, Si, Ca, and Ti abundances\footnote{If we exclude
Ti from the average as a result of its partial production in processes other than $\alpha$-captures \citep[e.g.,][]{Kobayashi2006}, the mean $\alpha$-enhancement increases merely by 0.02 dex.}, for NGC~1261, 
a sample of reference GCs that are a potential Gaia-Enceladus prodigy, and the stellar content of Gaia-Enceladus itself { (as contours)}.
\begin{figure}[htb]
\centering
\includegraphics[width=1\hsize]{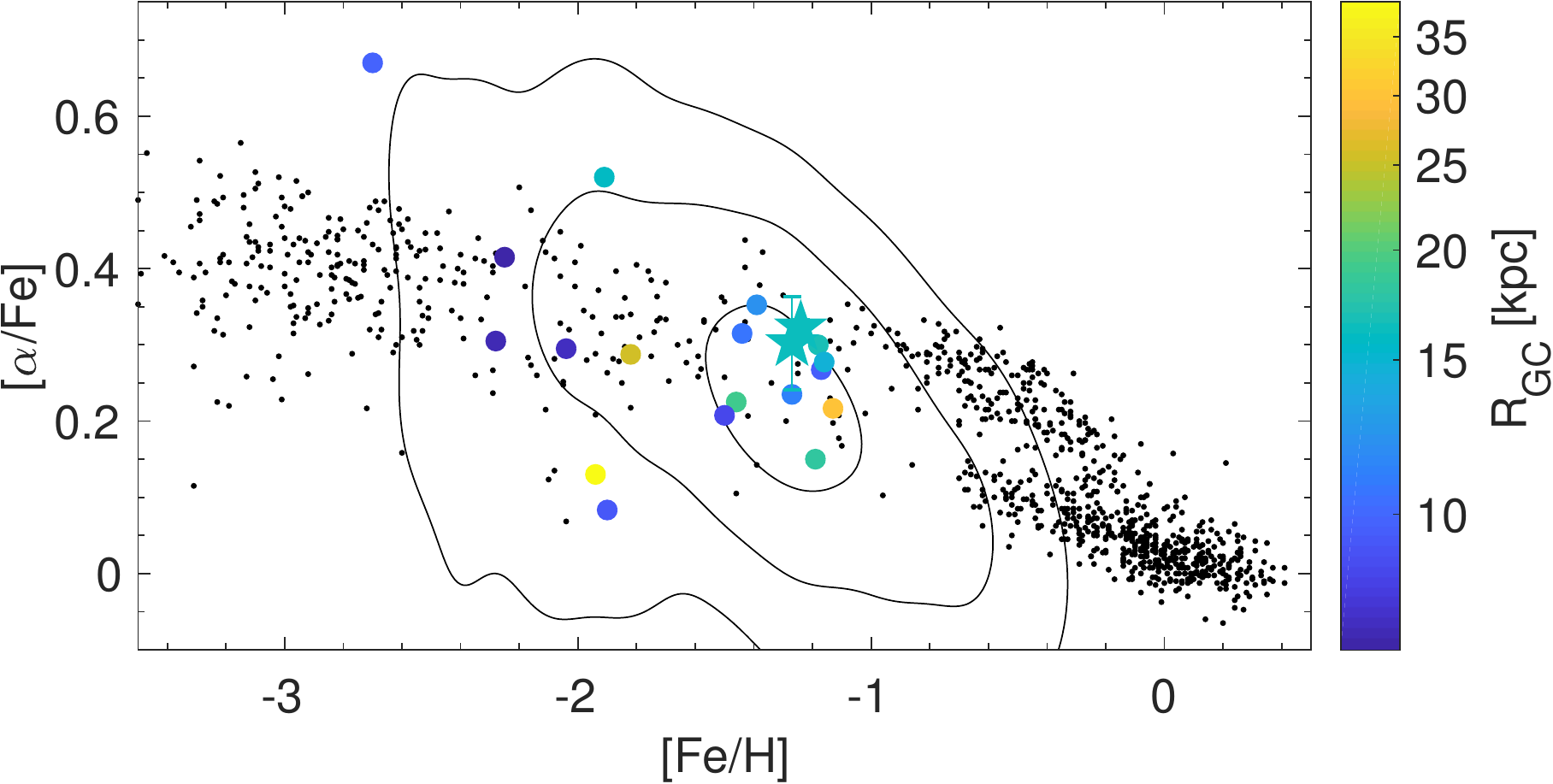}
\caption{Averaged $<$[Mg,Si,Ca,Ti/Fe]$>$ ratio for MW stars (black dots) and NGC~1261 (star symbols). 
{ The stellar component of Gaia-Enceladus  \citep{Helmi2018GaiaEnceladus} is overlaid as 1, 2, and 3$\sigma$ contours.
A sample of reference GCs dynamically associated with Gaia-Enceladus,  color-coded by their Galactocentric distance, are shown. The} 
GC abundances from the literature are from 
\citet{Francois1991},
\citet{McWilliam1992},
\citet{Shetrone2000},
\citet{Cohen2005},
\citet{Carretta2011},  
\citet{Roederer2011},
\citet{Kacharov2013},
\citet{Lovisi2013},
\citet{Khamidullina2014},
\citet{Koch5897},
\citet{Yong2014},
\citet{Roederer2015},
\citet{Carretta2013},  
\citet{Johnson2017},
\citet{Marino2017}, and
\citet{Koch2019Pal15}.
}
\end{figure}

Overall,  NGC~1261 appears chemically very similar to Gaia-Enceladus in that it occupies the  same mean metallicity and $\alpha$-enhancement, irrespective of the 
large abundance spread in the massive parent galaxy. 
\citet{Monty2020} report on the  $\alpha$-elements in Gaia-Enceladus and fit a toy model to their distribution with metallicity; in this work, the location of 
the knee, that is, the downturn in [$\alpha$/Fe] upon the onset of Fe-producing SNe Ia, lies at $\sim$$-$1.7 dex, placing it into the regime of the massive Galactic
satellite Fornax \citep{Hendricks2014,Reichert2020}. Accordingly, Gaia-Enceladus stars at the same
metallicity as NGC~1261 are 
expected to exhibit an [$\alpha$/Fe] ratio of $\sim$0.3 dex, which fully agrees with our measured values. 

The most important information can be gleaned from the heavy neutron-capture elements. Moderately $r$-process enriched r-I and r-II stars in the Galactic halo are often found to have chemodynamic associations amongst each other, 
arguing for an origin in common, disrupted objects
\citep[e.g.,][]{Roederer2018,Gudin2021}, although the fraction of bona fide accreted stars is not fully known due to the limited sample sizes. 
While similar enrichment mechanisms are likely to happen in the halo field 
the large fraction and preponderance of $r$-rich stars in faint dwarf galaxies is striking \citep[e.g.,][]{Ji2016,CJHansen2018,Reichert2021}.
Gaia-Enceladus also stands out in that it shows strong enhancement in the neutron-capture elements \citep{Aguado2021,Matsuno2021}.
In particular, the [Eu/Fe] ratios for the four stars in \citet[][see Fig.~7]{Aguado2021} are compatible with our findings in NGC~1261, albeit at slightly lower
metallicities of $-2.1$ to $-1.4$ dex, placing them in the border region of r-I and r-II stars.
By comparison with kinematically selected Gaia-Enceladus stars
from the Galah survey  \citep{Buder2020},  \citet{Aguado2021} highlight that the over-enhancements of 0.3$ <$ [Eu/Fe] $<$ 1 seen 
over the entire metallicity range of the progenitor (sampled over $\sim$1 dex) is a clear sign of $r$-process nucleosynthesis dominating the chemical evolution 
of this accreted satellite. This is clearly also seen in the associated NGC~1261, where similar enrichment took place before the GC was, still partly intact, accreted onto 
the MW. 

Exploring further a possible dwarf galaxy connection, we compare stars 35 and 46 to the recently discovered Eu-rich stars in Fornax \citep{Reichert2021}. 
These three stars have $-$1.3$<$[Fe/H]$<$$-$0.8 hence making for a reasonable small comparison sample in a more massive  satellite. 
The absolute Eu abundance of the three Eu stars in Fornax amount to almost 1 dex, which compares to a  $\log \varepsilon$\,(Eu) in NGC 1261 close to $-0.1$\,dex. 
Hence, the formation mechanism and/or mixing in Fornax and NGC 1261 must differ, despite the very similar [Fe/H] of the samples, and we expect both different formation mechanisms and mixing scenarios to be at work.  
Given the similarity of the Eu/Fe ratios in the two stars analyzed in this work, we may consider that the Eu enrichment is primordial and did not occur after the stars formed.
However, if the GC was enriched after one, or more, generations of stars formed, then stars below the turn-off must have even larger $r$-process abundances.
At the turn-off magnitude of V$\sim$19.5 mag, such a study is currently observationally too challenging to probe in statistically meaningful numbers.

Returning to the question of whether NGC~1261 could be associated with either the EriPhe overdensity or the { Phoenix} stellar stream, as 
advocated by  \citet{Shipp2018}, the need for detailed chemical tagging of halo substructures 
is undisputed \citep[e.g.,][]{Ji2020,TTHansen2020,Naidu2020,Prudil2021}.
In this context we note that 
\citet{Ji2020} find that their sample of eight { Phoenix} stream stars are very metal poor, at $<$$-2.6$ dex, and thus lie far below the metallicity we reported for our target cluster. 
Unless the {Phoenix} progenitor had a large mass allowing for an abundance spread of more than 1.5 dex \citep[cf.][]{Koch2009Review,Kirby2013}, 
we do not support any association with  NGC~1261.
\begin{acknowledgements}
We thank the anonymous referee for a positive and constructive report and are grateful to Moritz Reichert for helpful discussions.
AJKH gratefully acknowledges funding by the Deutsche Forschungsgemeinschaft (DFG, 
German Research Foundation) -- Project-ID 138713538 -- SFB 881 (``The Milky Way System''), subprojects A03, A05, A11. 
This project was developed in part at the Streams 21 meeting, virtually hosted by the Flatiron Institute.
 \end{acknowledgements}
\bibliographystyle{aa} 
\bibliography{ms} 

\begin{thebibliography}{153}
\expandafter\ifx\csname natexlab\endcsname\relax\def\natexlab#1{#1}\fi

\bibitem[{{Aguado} {et~al.}(2021){Aguado}, {Belokurov}, {Myeong}, {Evans},
  {Kobayashi}, {Sbordone}, {Chanam{\'e}}, {Navarrete}, \&
  {Koposov}}]{Aguado2021}
{Aguado}, D.~S., {Belokurov}, V., {Myeong}, G.~C., {et~al.} 2021, \apjl, 908,
  L8

\bibitem[{{Arellano Ferro} {et~al.}(2019){Arellano Ferro}, {Bustos Fierro},
  {Calder{\'o}n}, \& {Ahumada}}]{ArellanoFerro2019}
{Arellano Ferro}, A., {Bustos Fierro}, I.~H., {Calder{\'o}n}, J.~H., \&
  {Ahumada}, J.~A. 2019, \rmxaa, 55, 337

\bibitem[{{Arlandini} {et~al.}(1999){Arlandini}, {K{\"a}ppeler}, {Wisshak},
  {Gallino}, {Lugaro}, {Busso}, \& {Straniero}}]{Arlandini1999}
{Arlandini}, C., {K{\"a}ppeler}, F., {Wisshak}, K., {et~al.} 1999, \apj, 525,
  886

\bibitem[{{Asplund} {et~al.}(2009){Asplund}, {Grevesse}, {Sauval}, \&
  {Scott}}]{Asplund2009}
{Asplund}, M., {Grevesse}, N., {Sauval}, A.~J., \& {Scott}, P. 2009, \araa, 47,
  481

\bibitem[{{Balbinot} {et~al.}(2016){Balbinot}, {Yanny}, {Li}, {Santiago},
  {Marshall}, {Finley}, {Pieres}, {Abbott}, {Abdalla}, {Allam},
  {Benoit-L{\'e}vy}, {Bernstein}, {Bertin}, {Brooks}, {Burke}, {Carnero
  Rosell}, {Carrasco Kind}, {Carretero}, {Cunha}, {da Costa}, {DePoy}, {Desai},
  {Diehl}, {Doel}, {Estrada}, {Flaugher}, {Frieman}, {Gerdes}, {Gruen},
  {Gruendl}, {Honscheid}, {James}, {Kuehn}, {Kuropatkin}, {Lahav}, {March},
  {Martini}, {Miquel}, {Nichol}, {Ogando}, {Romer}, {Sanchez}, {Schubnell},
  {Sevilla-Noarbe}, {Smith}, {Soares-Santos}, {Sobreira}, {Suchyta}, {Tarle},
  {Thomas}, {Tucker}, {Walker}, \& {DES Collaboration}}]{Balbinot2016}
{Balbinot}, E., {Yanny}, B., {Li}, T.~S., {et~al.} 2016, \apj, 820, 58

\bibitem[{{Barbuy} {et~al.}(2015){Barbuy}, {Fria{\c{c}}a}, {da Silveira},
  {Hill}, {Zoccali}, {Minniti}, {Renzini}, {Ortolani}, \&
  {G{\'o}mez}}]{Barbuy2015}
{Barbuy}, B., {Fria{\c{c}}a}, A.~C.~S., {da Silveira}, C.~R., {et~al.} 2015,
  \aap, 580, A40

\bibitem[{{Bastian} \& {Lardo}(2018)}]{Bastian2018}
{Bastian}, N. \& {Lardo}, C. 2018, \araa, 56, 83

\bibitem[{{Battistini} \& {Bensby}(2015)}]{Battistini2015}
{Battistini}, C. \& {Bensby}, T. 2015, \aap, 577, A9

\bibitem[{{Battistini} \& {Bensby}(2016)}]{Battistini2016}
{Battistini}, C. \& {Bensby}, T. 2016, \aap, 586, A49

\bibitem[{{Baumgardt} {et~al.}(2019){Baumgardt}, {Hilker}, {Sollima}, \&
  {Bellini}}]{Baumgardt2019}
{Baumgardt}, H., {Hilker}, M., {Sollima}, A., \& {Bellini}, A. 2019, \mnras,
  482, 5138

\bibitem[{{Beers} \& {Christlieb}(2005)}]{BeersChristlieb2005}
{Beers}, T.~C. \& {Christlieb}, N. 2005, \araa, 43, 531

\bibitem[{{Belokurov} {et~al.}(2018){Belokurov}, {Erkal}, {Evans}, {Koposov},
  \& {Deason}}]{Belokurov2018}
{Belokurov}, V., {Erkal}, D., {Evans}, N.~W., {Koposov}, S.~E., \& {Deason},
  A.~J. 2018, \mnras, 478, 611

\bibitem[{{Bensby} {et~al.}(2014){Bensby}, {Feltzing}, \& {Oey}}]{Bensby2014}
{Bensby}, T., {Feltzing}, S., \& {Oey}, M.~S. 2014, \aap, 562, A71

\bibitem[{{Bergemann} {et~al.}(2012){Bergemann}, {Hansen}, {Bautista}, \&
  {Ruchti}}]{Bergemann2012}
{Bergemann}, M., {Hansen}, C.~J., {Bautista}, M., \& {Ruchti}, G. 2012, \aap,
  546, A90

\bibitem[{{Bisterzo} {et~al.}(2012){Bisterzo}, {Gallino}, {Straniero},
  {Cristallo}, \& {K{\"a}ppeler}}]{Bisterzo2012}
{Bisterzo}, S., {Gallino}, R., {Straniero}, O., {Cristallo}, S., \&
  {K{\"a}ppeler}, F. 2012, \mnras, 422, 849

\bibitem[{{Buder} {et~al.}(2021){Buder}, {Sharma}, {Kos}, {Amarsi},
  {Nordlander}, {Lind}, {Martell}, {Asplund}, {Bland-Hawthorn}, {Casey}, {de
  Silva}, {D'Orazi}, {Freeman}, {Hayden}, {Lewis}, {Lin}, {Schlesinger},
  {Simpson}, {Stello}, {Zucker}, {Zwitter}, {Beeson}, {Buck}, {Casagrande},
  {Clark}, {{\v{C}}otar}, {da Costa}, {de Grijs}, {Feuillet}, {Horner},
  {Kafle}, {Khanna}, {Kobayashi}, {Liu}, {Montet}, {Nandakumar}, {Nataf},
  {Ness}, {Spina}, {Tepper-Garc{\'\i}a}, {Ting}, {Traven},
  {Vogrin{\v{c}}i{\v{c}}}, {Wittenmyer}, {Wyse}, {{\v{Z}}erjal}, \& {Galah
  Collaboration}}]{Buder2020}
{Buder}, S., {Sharma}, S., {Kos}, J., {et~al.} 2021, \mnras

\bibitem[{{Burris} {et~al.}(2000){Burris}, {Pilachowski}, {Armand roff},
  {Sneden}, {Cowan}, \& {Roe}}]{Burris2000}
{Burris}, D.~L., {Pilachowski}, C.~A., {Armand roff}, T.~E., {et~al.} 2000,
  \apj, 544, 302

\bibitem[{{Busso} {et~al.}(1999){Busso}, {Gallino}, \&
  {Wasserburg}}]{Busso1999}
{Busso}, M., {Gallino}, R., \& {Wasserburg}, G.~J. 1999, \araa, 37, 239

\bibitem[{{Carretta} {et~al.}(2009){Carretta}, {Bragaglia}, {Gratton}, \&
  {Lucatello}}]{Carretta2009NaO}
{Carretta}, E., {Bragaglia}, A., {Gratton}, R., \& {Lucatello}, S. 2009, \aap,
  505, 139

\bibitem[{{Carretta} {et~al.}(2011){Carretta}, {Lucatello}, {Gratton},
  {Bragaglia}, \& {D'Orazi}}]{Carretta2011}
{Carretta}, E., {Lucatello}, S., {Gratton}, R.~G., {Bragaglia}, A., \&
  {D'Orazi}, V. 2011, \aap, 533, A69

\bibitem[{{Carretta} {et~al.}(2012){Carretta}, {D'Orazi}, {Gratton}, \&
  {Lucatello}}]{Carretta2012}
{Carretta}, E., {D'Orazi}, V., {Gratton}, R.~G., \& {Lucatello}, S. 2012, \aap,
  543, A117

\bibitem[{{Carretta} {et~al.}(2013){Carretta}, {Bragaglia}, {Gratton},
  {Lucatello}, {D'Orazi}, {Bellazzini}, {Catanzaro}, {Leone}, {Momany}, \&
  {Sollima}}]{Carretta2013}
{Carretta}, E., {Bragaglia}, A., {Gratton}, R.~G., {et~al.} 2013, \aap, 557,
  A138

\bibitem[{{Castelli} \& {Kurucz}(2003)}]{CastelliKurucz2003}
{Castelli}, F. \& {Kurucz}, R.~L. 2003, in IAU Symposium, Vol. 210, Modelling
  of Stellar Atmospheres, ed. N.~{Piskunov}, W.~W. {Weiss}, \& D.~F. {Gray},
  A20

\bibitem[{{Choplin} {et~al.}(2018){Choplin}, {Hirschi}, {Meynet},
  {Ekstr{\"o}m}, {Chiappini}, \& {Laird}}]{Choplin2018}
{Choplin}, A., {Hirschi}, R., {Meynet}, G., {et~al.} 2018, \aap, 618, A133

\bibitem[{{Cohen}(1978)}]{Cohen1978}
{Cohen}, J.~G. 1978, \apj, 223, 487

\bibitem[{{Cohen} \& {Melendez}(2005)}]{Cohen2005}
{Cohen}, J.~G. \& {Melendez}, J. 2005, \aj, 129, 1607

\bibitem[{{Cowan} \& {Rose}(1977)}]{CowanRose1977}
{Cowan}, J.~J. \& {Rose}, W.~K. 1977, \apj, 212, 149

\bibitem[{{Cristallo} {et~al.}(2011){Cristallo}, {Piersanti}, {Straniero},
  {Gallino}, {Dom{\'{\i}}nguez}, {Abia}, {Di Rico}, {Quintini}, \&
  {Bisterzo}}]{Cristallo2011}
{Cristallo}, S., {Piersanti}, L., {Straniero}, O., {et~al.} 2011, \apjs, 197,
  17

\bibitem[{{Cristallo} {et~al.}(2015{\natexlab{a}}){Cristallo}, {Abia},
  {Straniero}, \& {Piersanti}}]{Cristallo2015}
{Cristallo}, S., {Abia}, C., {Straniero}, O., \& {Piersanti}, L.
  2015{\natexlab{a}}, \apj, 801, 53

\bibitem[{{Cristallo} {et~al.}(2015{\natexlab{b}}){Cristallo}, {Straniero},
  {Piersanti}, \& {Gobrecht}}]{Cristallo2015fruity}
{Cristallo}, S., {Straniero}, O., {Piersanti}, L., \& {Gobrecht}, D.
  2015{\natexlab{b}}, \apjs, 219, 40

\bibitem[{{Den Hartog} {et~al.}(2003){Den Hartog}, {Lawler}, {Sneden}, \&
  {Cowan}}]{DenHartog2003}
{Den Hartog}, E.~A., {Lawler}, J.~E., {Sneden}, C., \& {Cowan}, J.~J. 2003,
  \apjs, 148, 543

\bibitem[{{Den Hartog} {et~al.}(2011){Den Hartog}, {Lawler}, {Sobeck},
  {Sneden}, \& {Cowan}}]{DenHartog2011}
{Den Hartog}, E.~A., {Lawler}, J.~E., {Sobeck}, J.~S., {Sneden}, C., \&
  {Cowan}, J.~J. 2011, \apjs, 194, 35

\bibitem[{{Ferraro} {et~al.}(1999){Ferraro}, {Messineo}, {Fusi Pecci}, {de
  Palo}, {Straniero}, {Chieffi}, \& {Limongi}}]{Ferraro1999}
{Ferraro}, F.~R., {Messineo}, M., {Fusi Pecci}, F., {et~al.} 1999, \aj, 118,
  1738

\bibitem[{{Filler} {et~al.}(2012){Filler}, {Ivans}, \& {Simmerer}}]{Filler2012}
{Filler}, D., {Ivans}, I.~I., \& {Simmerer}, J. 2012, in American Astronomical
  Society Meeting Abstracts, Vol. 219, American Astronomical Society Meeting
  Abstracts \#219, 152.05

\bibitem[{{Fran{\c{c}}ois} {et~al.}(2007){Fran{\c{c}}ois}, {Depagne}, {Hill},
  {Spite}, {Spite}, {Plez}, {Beers}, {Andersen}, {James}, {Barbuy}, {Cayrel},
  {Bonifacio}, {Molaro}, {Nordstr{\"o}m}, \& {Primas}}]{Francois2007}
{Fran{\c{c}}ois}, P., {Depagne}, E., {Hill}, V., {et~al.} 2007, \aap, 476, 935

\bibitem[{{Francois}(1991)}]{Francois1991}
{Francois}, P. 1991, \aap, 247, 56

\bibitem[{{Frischknecht} {et~al.}(2012){Frischknecht}, {Hirschi}, \&
  {Thielemann}}]{Frischknecht2012}
{Frischknecht}, U., {Hirschi}, R., \& {Thielemann}, F.-K. 2012, \aap, 538, L2

\bibitem[{{Frischknecht} {et~al.}(2016){Frischknecht}, {Hirschi}, {Pignatari},
  {Maeder}, {Meynet}, {Chiappini}, {Thielemann}, {Rauscher}, {Georgy}, \&
  {Ekstr{\"o}m}}]{Frischknecht2016}
{Frischknecht}, U., {Hirschi}, R., {Pignatari}, M., {et~al.} 2016, \mnras, 456,
  1803

\bibitem[{{Gaia Collaboration} {et~al.}(2018){Gaia Collaboration}, {Brown},
  {Vallenari}, {Prusti}, {de Bruijne}, {Babusiaux}, {Bailer-Jones}, {Biermann},
  {Evans}, {Eyer}, \& et~al.}]{GaiaDR2}
{Gaia Collaboration}, {Brown}, A.~G.~A., {Vallenari}, A., {et~al.} 2018, \aap,
  616, A1

\bibitem[{{Gnedin} \& {Ostriker}(1997)}]{Gnedin1997}
{Gnedin}, O.~Y. \& {Ostriker}, J.~P. 1997, \apj, 474, 223

\bibitem[{{Goriely} {et~al.}(2011){Goriely}, {Bauswein}, \&
  {Janka}}]{Goriely2011}
{Goriely}, S., {Bauswein}, A., \& {Janka}, H.-T. 2011, \apjl, 738, L32

\bibitem[{{Gudin} {et~al.}(2021){Gudin}, {Shank}, {Beers}, {Yuan}, {Limberg},
  {Roederer}, {Placco}, {Holmbeck}, {Dietz}, {Rasmussen}, {Hansen}, {Sakari},
  {Ezzeddine}, \& {Frebel}}]{Gudin2021}
{Gudin}, D., {Shank}, D., {Beers}, T.~C., {et~al.} 2021, \apj, 908, 79

\bibitem[{{Hampel} {et~al.}(2016){Hampel}, {Stancliffe}, {Lugaro}, \&
  {Meyer}}]{Hampel2016}
{Hampel}, M., {Stancliffe}, R.~J., {Lugaro}, M., \& {Meyer}, B.~S. 2016, \apj,
  831, 171

\bibitem[{{Hanke} {et~al.}(2020{\natexlab{a}}){Hanke}, {Hansen}, {Ludwig},
  {Cristallo}, {McWilliam}, {Grebel}, \& {Piersanti}}]{Hanke2020HD20}
{Hanke}, M., {Hansen}, C.~J., {Ludwig}, H.-G., {et~al.} 2020{\natexlab{a}},
  \aap, 635, A104

\bibitem[{{Hanke} {et~al.}(2020{\natexlab{b}}){Hanke}, {Koch}, {Prudil},
  {Grebel}, \& {Bastian}}]{Hanke2020}
{Hanke}, M., {Koch}, A., {Prudil}, Z., {Grebel}, E.~K., \& {Bastian}, U.
  2020{\natexlab{b}}, \aap, 637, A98

\bibitem[{{Hansen} {et~al.}(2012){Hansen}, {Primas}, {Hartman}, {Kratz},
  {Wanajo}, {Leibundgut}, {Farouqi}, {Hallmann}, {Christlieb}, \&
  {Nilsson}}]{CJHansen2012}
{Hansen}, C.~J., {Primas}, F., {Hartman}, H., {et~al.} 2012, \aap, 545, A31

\bibitem[{{Hansen} {et~al.}(2013){Hansen}, {Bergemann}, {Cescutti}, {Fran{\c
  c}ois}, {Arcones}, {Karakas}, {Lind}, \& {Chiappini}}]{CJHansen2013}
{Hansen}, C.~J., {Bergemann}, M., {Cescutti}, G., {et~al.} 2013, \aap, 551, A57

\bibitem[{{Hansen} {et~al.}(2014){Hansen}, {Montes}, \&
  {Arcones}}]{CJHansen2014}
{Hansen}, C.~J., {Montes}, F., \& {Arcones}, A. 2014, \apj, 797, 123

\bibitem[{{Hansen} {et~al.}(2018){Hansen}, {El-Souri}, {Monaco}, {Villanova},
  {Bonifacio}, {Caffau}, \& {Sbordone}}]{CJHansen2018}
{Hansen}, C.~J., {El-Souri}, M., {Monaco}, L., {et~al.} 2018, \apj, 855, 83

\bibitem[{{Hansen} {et~al.}(2020){Hansen}, {Riley}, {Strigari}, {Marshall},
  {Ferguson}, {Zepeda}, \& {Sneden}}]{TTHansen2020}
{Hansen}, T.~T., {Riley}, A.~H., {Strigari}, L.~E., {et~al.} 2020, \apj, 901,
  23

\bibitem[{{Harris}(1996)}]{Harris1996}
{Harris}, W.~E. 1996, \aj, 112, 1487

\bibitem[{{Helmi} {et~al.}(2018){Helmi}, {Babusiaux}, {Koppelman}, {Massari},
  {Veljanoski}, \& {Brown}}]{Helmi2018GaiaEnceladus}
{Helmi}, A., {Babusiaux}, C., {Koppelman}, H.~H., {et~al.} 2018, \nat, 563, 85

\bibitem[{{Hendricks} {et~al.}(2014){Hendricks}, {Koch}, {Lanfranchi},
  {Boeche}, {Walker}, {Johnson}, {Pe{\~n}arrubia}, \&
  {Gilmore}}]{Hendricks2014}
{Hendricks}, B., {Koch}, A., {Lanfranchi}, G.~A., {et~al.} 2014, \apj, 785, 102

\bibitem[{{Hidalgo} {et~al.}(2018){Hidalgo}, {Pietrinferni}, {Cassisi},
  {Salaris}, {Mucciarelli}, {Savino}, {Aparicio}, {Silva Aguirre}, \&
  {Verma}}]{Hidalgo2018}
{Hidalgo}, S.~L., {Pietrinferni}, A., {Cassisi}, S., {et~al.} 2018, \apj, 856,
  125

\bibitem[{{Hirschi}(2007)}]{Hirschi2007}
{Hirschi}, R. 2007, \aap, 461, 571

\bibitem[{{Horta} {et~al.}(2020){Horta}, {Schiavon}, {Mackereth}, {Beers},
  {Fern{\'a}ndez-Trincado}, {Frinchaboy}, {Garc{\'\i}a-Hern{\'a}ndez},
  {Geisler}, {Hasselquist}, {J{\"o}nsson}, {Lane}, {Majewski},
  {M{\'e}sz{\'a}ros}, {Bidin}, {Nataf}, {Roman-Lopes}, {Nitschelm},
  {Vargas-Gonz{\'a}lez}, \& {Zasowski}}]{Horta2020}
{Horta}, D., {Schiavon}, R.~P., {Mackereth}, J.~T., {et~al.} 2020, \mnras, 493,
  3363

\bibitem[{{Horta} {et~al.}(2021){Horta}, {Schiavon}, {Mackereth}, {Pfeffer},
  {Mason}, {Kisku}, {Fragkoudi}, {Allende Prieto}, {Cunha}, {Hasselquist},
  {Holtzman}, {Majewski}, {Nataf}, {O'Connell}, {Schultheis}, \&
  {Smith}}]{Horta2021}
{Horta}, D., {Schiavon}, R.~P., {Mackereth}, J.~T., {et~al.} 2021, \mnras, 500,
  1385

\bibitem[{{Ji} {et~al.}(2016){Ji}, {Frebel}, {Simon}, \& {Chiti}}]{Ji2016}
{Ji}, A.~P., {Frebel}, A., {Simon}, J.~D., \& {Chiti}, A. 2016, \apj, 830, 93

\bibitem[{{Ji} {et~al.}(2020){Ji}, {Li}, {Hansen}, {Casey}, {Koposov}, {Pace},
  {Mackey}, {Lewis}, {Simpson}, {Bland-Hawthorn}, {Cullinane}, {Da Costa},
  {Hattori}, {Martell}, {Kuehn}, {Erkal}, {Shipp}, {Wan}, \& {Zucker}}]{Ji2020}
{Ji}, A.~P., {Li}, T.~S., {Hansen}, T.~T., {et~al.} 2020, \aj, 160, 181

\bibitem[{{Johnson} {et~al.}(2013){Johnson}, {Rich}, {Kobayashi}, {Kunder},
  {Pilachowski}, {Koch}, \& {de Propris}}]{Johnson2013}
{Johnson}, C.~I., {Rich}, R.~M., {Kobayashi}, C., {et~al.} 2013, \apj, 765, 157

\bibitem[{{Johnson} {et~al.}(2017){Johnson}, {Caldwell}, {Rich}, \&
  {Walker}}]{Johnson2017}
{Johnson}, C.~I., {Caldwell}, N., {Rich}, R.~M., \& {Walker}, M.~G. 2017, \aj,
  154, 155

\bibitem[{{Kacharov} {et~al.}(2013){Kacharov}, {Koch}, \&
  {McWilliam}}]{Kacharov2013}
{Kacharov}, N., {Koch}, A., \& {McWilliam}, A. 2013, \aap, 554, A81

\bibitem[{{Khamidullina} {et~al.}(2014){Khamidullina}, {Sharina}, {Shimansky},
  \& {Davoust}}]{Khamidullina2014}
{Khamidullina}, D.~A., {Sharina}, M.~E., {Shimansky}, V.~V., \& {Davoust}, E.
  2014, Astrophysical Bulletin, 69, 409

\bibitem[{{Kirby} {et~al.}(2013){Kirby}, {Cohen}, {Guhathakurta}, {Cheng},
  {Bullock}, \& {Gallazzi}}]{Kirby2013}
{Kirby}, E.~N., {Cohen}, J.~G., {Guhathakurta}, P., {et~al.} 2013, \apj, 779,
  102

\bibitem[{{Kirby} {et~al.}(2015){Kirby}, {Guo}, {Zhang}, {Deng}, {Cohen},
  {Guhathakurta}, {Shetrone}, {Lee}, \& {Rizzi}}]{Kirby2015}
{Kirby}, E.~N., {Guo}, M., {Zhang}, A.~J., {et~al.} 2015, \apj, 801, 125

\bibitem[{{Kobayashi} {et~al.}(2006){Kobayashi}, {Umeda}, {Nomoto}, {Tominaga},
  \& {Ohkubo}}]{Kobayashi2006}
{Kobayashi}, C., {Umeda}, H., {Nomoto}, K., {Tominaga}, N., \& {Ohkubo}, T.
  2006, \apj, 653, 1145

\bibitem[{{Koch}(2009)}]{Koch2009Review}
{Koch}, A. 2009, Astronomische Nachrichten, 330, 675

\bibitem[{{Koch} \& {McWilliam}(2008)}]{Koch2008}
{Koch}, A. \& {McWilliam}, A. 2008, \aj, 135, 1551

\bibitem[{{Koch} {et~al.}(2008){Koch}, {Grebel}, {Gilmore}, {Wyse}, {Kleyna},
  {Harbeck}, {Wilkinson}, \& {Evans}}]{Koch2008Carina}
{Koch}, A., {Grebel}, E.~K., {Gilmore}, G.~F., {et~al.} 2008, \aj, 135, 1580

\bibitem[{{Koch} \& {McWilliam}(2014)}]{Koch5897}
{Koch}, A. \& {McWilliam}, A. 2014, \aap, 565, A23

\bibitem[{{Koch} \& {C{\^o}t{\'e}}(2019)}]{Koch2019Pal13}
{Koch}, A. \& {C{\^o}t{\'e}}, P. 2019, \aap, 632, A55

\bibitem[{{Koch} {et~al.}(2019{\natexlab{a}}){Koch}, {Grebel}, \&
  {Martell}}]{Koch2019CN}
{Koch}, A., {Grebel}, E.~K., \& {Martell}, S.~L. 2019{\natexlab{a}}, \aap, 625,
  A75

\bibitem[{{Koch} {et~al.}(2019{\natexlab{b}}){Koch}, Xi, \&
  Rich}]{Koch2019Pal15}
{Koch}, A., Xi, S., \& Rich, R. 2019{\natexlab{b}}, \aap, 627, A70

\bibitem[{{Kraft} \& {Ivans}(2003)}]{KraftIvans2003}
{Kraft}, R.~P. \& {Ivans}, I.~I. 2003, \pasp, 115, 143

\bibitem[{{Kravtsov} {et~al.}(2010){Kravtsov}, {Alca{\'\i}no}, {Marconi}, \&
  {Alvarado}}]{Kravtsov2010}
{Kravtsov}, V., {Alca{\'\i}no}, G., {Marconi}, G., \& {Alvarado}, F. 2010,
  \aap, 516, A23

\bibitem[{{Kruijssen} {et~al.}(2019){Kruijssen}, {Pfeffer}, {Reina-Campos},
  {Crain}, \& {Bastian}}]{Kruijssen2019}
{Kruijssen}, J.~M.~D., {Pfeffer}, J.~L., {Reina-Campos}, M., {Crain}, R.~A., \&
  {Bastian}, N. 2019, \mnras, 486, 3180

\bibitem[{{Kuzma} {et~al.}(2018){Kuzma}, {Da Costa}, \& {Mackey}}]{Kuzma2018}
{Kuzma}, P.~B., {Da Costa}, G.~S., \& {Mackey}, A.~D. 2018, \mnras, 473, 2881

\bibitem[{{Lawler} {et~al.}(2006){Lawler}, {Den Hartog}, {Sneden}, \&
  {Cowan}}]{Lawler2006}
{Lawler}, J.~E., {Den Hartog}, E.~A., {Sneden}, C., \& {Cowan}, J.~J. 2006,
  \apjs, 162, 227


\bibitem[{{Lawler} {et~al.}(2009){Lawler}, {Sneden}, {Cowan}, {Ivans}, \& {Den
  Hartog}}]{Lawler2009}
{Lawler}, J.~E., {Sneden}, C., {Cowan}, J.~J., {Ivans}, I.~I., \& {Den Hartog},
  E.~A. 2009, \apjs, 182, 51

\bibitem[{{Lawler} {et~al.}(2015){Lawler}, {Sneden}, \& {Cowan}}]{Lawler2015}
{Lawler}, J.~E., {Sneden}, C., \& {Cowan}, J.~J. 2015, \apjs, 220, 13

\bibitem[{{Lawler} {et~al.}(2019){Lawler}, {Hala}, {Sneden}, {Nave}, {Wood}, \&
  {Cowan}}]{Lawler2019}
{Lawler}, J.~E., {Hala}, {Sneden}, C., {et~al.} 2019, \apjs, 241, 21

\bibitem[{{Lawler} {et~al.}(2001){Lawler}, {Wickliffe}, {den Hartog}, \&
  {Sneden}}]{Lawler2001}
{Lawler}, J.~E., {Wickliffe}, M.~E., {den Hartog}, E.~A., \& {Sneden}, C. 2001,
  \apj, 563, 1075

\bibitem[{{Lee} {et~al.}(1999){Lee}, {Joo}, {Sohn}, {Rey}, {Lee}, \&
  {Walker}}]{Lee1999}
{Lee}, Y.~W., {Joo}, J.~M., {Sohn}, Y.~J., {et~al.} 1999, \nat, 402, 55

\bibitem[{{Leon} {et~al.}(2000){Leon}, {Meylan}, \& {Combes}}]{Leon2000}
{Leon}, S., {Meylan}, G., \& {Combes}, F. 2000, \aap, 359, 907

\bibitem[{{Li} {et~al.}(2016){Li}, {Balbinot}, {Mondrik}, {Marshall}, {Yanny},
  {Bechtol}, {Drlica-Wagner}, {Oscar}, {Santiago}, {Simon}, {Vivas}, {Walker},
  {Wang}, {Abbott}, {Abdalla}, {Benoit-L{\'e}vy}, {Bernstein}, {Bertin},
  {Brooks}, {Burke}, {Carnero Rosell}, {Carrasco Kind}, {Carretero}, {da
  Costa}, {DePoy}, {Desai}, {Diehl}, {Doel}, {Estrada}, {Finley}, {Flaugher},
  {Frieman}, {Gruen}, {Gruendl}, {Gutierrez}, {Honscheid}, {James}, {Kuehn},
  {Kuropatkin}, {Lahav}, {Maia}, {March}, {Martini}, {Ogando}, {Plazas},
  {Reil}, {Romer}, {Roodman}, {Sanchez}, {Scarpine}, {Schubnell},
  {Sevilla-Noarbe}, {Smith}, {Soares-Santos}, {Sobreira}, {Suchyta}, {Swanson},
  {Tarle}, {Tucker}, {Zhang}, \& {DES Collaboration}}]{Li2016}
{Li}, T.~S., {Balbinot}, E., {Mondrik}, N., {et~al.} 2016, \apj, 817, 135

\bibitem[{{Lind} {et~al.}(2009){Lind}, {Primas}, {Charbonnel}, {Grundahl}, \&
  {Asplund}}]{Lind2009}
{Lind}, K., {Primas}, F., {Charbonnel}, C., {Grundahl}, F., \& {Asplund}, M.
  2009, \aap, 503, 545

\bibitem[{{Lind} {et~al.}(2011){Lind}, {Charbonnel}, {Decressin}, {Primas},
  {Grundahl}, \& {Asplund}}]{Lind2011NGC6397}
{Lind}, K., {Charbonnel}, C., {Decressin}, T., {et~al.} 2011, \aap, 527, A148

\bibitem[{{Lovisi} {et~al.}(2013){Lovisi}, {Mucciarelli}, {Lanzoni}, {Ferraro},
  {Dalessandro}, \& {Monaco}}]{Lovisi2013}
{Lovisi}, L., {Mucciarelli}, A., {Lanzoni}, B., {et~al.} 2013, \apj, 772, 148

\bibitem[{{Malhan} {et~al.}(2019){Malhan}, {Ibata}, {Carlberg}, {Valluri}, \&
  {Freese}}]{Malhan2019}
{Malhan}, K., {Ibata}, R.~A., {Carlberg}, R.~G., {Valluri}, M., \& {Freese}, K.
  2019, \apj, 881, 106

\bibitem[{{Mar{\'{\i}}n-Franch} {et~al.}(2009){Mar{\'{\i}}n-Franch},
  {Aparicio}, {Piotto}, {Rosenberg}, {Chaboyer}, {Sarajedini}, {Siegel},
  {Anderson}, {Bedin}, {Dotter}, {Hempel}, {King}, {Majewski}, {Milone},
  {Paust}, \& {Reid}}]{Marin-Franch2009}
{Mar{\'{\i}}n-Franch}, A., {Aparicio}, A., {Piotto}, G., {et~al.} 2009, \apj,
  694, 1498

\bibitem[{{Marino} {et~al.}(2017){Marino}, {Milone}, {Yong}, {Da Costa},
  {Asplund}, {Bedin}, {Jerjen}, {Nardiello}, {Piotto}, \&
  {Renzini}}]{Marino2017}
{Marino}, A.~F., {Milone}, A.~P., {Yong}, D., {et~al.} 2017, \apj, 843, 66

\bibitem[{{Martell} \& {Grebel}(2010)}]{Martell2010}
{Martell}, S.~L. \& {Grebel}, E.~K. 2010, \aap, 519, A14

\bibitem[{{Mashonkina} \& {Christlieb}(2014)}]{Mashonkina2014}
{Mashonkina}, L. \& {Christlieb}, N. 2014, \aap, 565, A123

\bibitem[{{Massari} {et~al.}(2019){Massari}, {Koppelman}, \&
  {Helmi}}]{Massari2019}
{Massari}, D., {Koppelman}, H.~H., \& {Helmi}, A. 2019, \aap, 630, L4

\bibitem[{{Matsuno} {et~al.}(2021){Matsuno}, {Hirai}, {Tarumi}, {Hotokezaka},
  {Tanaka}, \& {Helmi}}]{Matsuno2021}
{Matsuno}, T., {Hirai}, Y., {Tarumi}, Y., {et~al.} 2021, \aap, 650, A110

\bibitem[{{Matteucci} \& {Brocato}(1990)}]{Matteucci1990}
{Matteucci}, F. \& {Brocato}, E. 1990, \apj, 365, 539

\bibitem[{{McWilliam} {et~al.}(1992){McWilliam}, {Geisler}, \&
  {Rich}}]{McWilliam1992}
{McWilliam}, A., {Geisler}, D., \& {Rich}, R.~M. 1992, \pasp, 104, 1193

\bibitem[{{McWilliam} {et~al.}(1995){McWilliam}, {Preston}, {Sneden}, \&
  {Searle}}]{McWilliam1995}
{McWilliam}, A., {Preston}, G.~W., {Sneden}, C., \& {Searle}, L. 1995, \aj,
  109, 2757

\bibitem[{{McWilliam}(1998)}]{McWilliam1998}
{McWilliam}, A. 1998, \aj, 115, 1640

\bibitem[{{Mishenina} {et~al.}(2002){Mishenina}, {Kovtyukh}, {Soubiran},
  {Travaglio}, \& {Busso}}]{Mishenina2002}
{Mishenina}, T.~V., {Kovtyukh}, V.~V., {Soubiran}, C., {Travaglio}, C., \&
  {Busso}, M. 2002, \aap, 396, 189

\bibitem[{{Mishenina} {et~al.}(2011){Mishenina}, {Gorbaneva}, {Basak},
  {Soubiran}, \& {Kovtyukh}}]{Mishenina2011}
{Mishenina}, T.~V., {Gorbaneva}, T.~I., {Basak}, N.~Y., {Soubiran}, C., \&
  {Kovtyukh}, V.~V. 2011, Astronomy Reports, 55, 689

\bibitem[{{Molero} {et~al.}(2021){Molero}, {Simonetti}, {Matteucci}, \& {della
  Valle}}]{Molero2021}
{Molero}, M., {Simonetti}, P., {Matteucci}, F., \& {della Valle}, M. 2021,
  \mnras, 500, 1071

\bibitem[{{Monty} {et~al.}(2020){Monty}, {Venn}, {Lane}, {Lokhorst}, \&
  {Yong}}]{Monty2020}
{Monty}, S., {Venn}, K.~A., {Lane}, J. M.~M., {Lokhorst}, D., \& {Yong}, D.
  2020, \mnras, 497, 1236

\bibitem[{{Mucciarelli} \& {Bonifacio}(2020)}]{Mucciarelli2020}
{Mucciarelli}, A. \& {Bonifacio}, P. 2020, \aap, 640, A87

\bibitem[{{Myeong} {et~al.}(2019){Myeong}, {Vasiliev}, {Iorio}, {Evans}, \&
  {Belokurov}}]{Myeong2019}
{Myeong}, G.~C., {Vasiliev}, E., {Iorio}, G., {Evans}, N.~W., \& {Belokurov},
  V. 2019, \mnras, 1731

\bibitem[{{Naidu} {et~al.}(2020){Naidu}, {Conroy}, {Bonaca}, {Johnson}, {Ting},
  {Caldwell}, {Zaritsky}, \& {Cargile}}]{Naidu2020}
{Naidu}, R.~P., {Conroy}, C., {Bonaca}, A., {et~al.} 2020, \apj, 901, 48

\bibitem[{{Naidu} {et~al.}(2021){Naidu}, {Conroy}, {Bonaca}, {Zaritsky},
  {Weinberger}, {Ting}, {Caldwell}, {Tacchella}, {Han}, \&
  {Speagle}}]{Naidu2021}
{Naidu}, R.~P., {Conroy}, C., {Bonaca}, A., {et~al.} 2021, arXiv e-prints,
  arXiv:2103.03251

\bibitem[{{Nissen} \& {Schuster}(1997)}]{Nissen1997}
{Nissen}, P.~E. \& {Schuster}, W.~J. 1997, \aap, 326, 751

\bibitem[{{Nissen} \& {Schuster}(2010)}]{NissenSchuster2010}
{Nissen}, P.~E. \& {Schuster}, W.~J. 2010, \aap, 511, L10

\bibitem[{{Odenkirchen} {et~al.}(2001){Odenkirchen}, {Grebel}, {Rockosi},
  {Dehnen}, {Ibata}, {Rix}, {Stolte}, {Wolf}, {Anderson}, {Bahcall},
  {Brinkmann}, {Csabai}, {Hennessy}, {Hindsley}, {Ivezi{\'c}}, {Lupton},
  {Munn}, {Pier}, {Stoughton}, \& {York}}]{Odenkirchen2001}
{Odenkirchen}, M., {Grebel}, E.~K., {Rockosi}, C.~M., {et~al.} 2001, \apjl,
  548, L165

\bibitem[{{Otsuki} {et~al.}(2006){Otsuki}, {Honda}, {Aoki}, {Kajino}, \&
  {Mathews}}]{Otsuki2006}
{Otsuki}, K., {Honda}, S., {Aoki}, W., {Kajino}, T., \& {Mathews}, G.~J. 2006,
  \apjl, 641, L117

\bibitem[{{Piersanti} {et~al.}(2013){Piersanti}, {Cristallo}, \&
  {Straniero}}]{Piersanti2013}
{Piersanti}, L., {Cristallo}, S., \& {Straniero}, O. 2013, \apj, 774, 98

\bibitem[{{Pignatari} {et~al.}(2010){Pignatari}, {Gallino}, {Heil}, {Wiescher},
  {K{\"a}ppeler}, {Herwig}, \& {Bisterzo}}]{Pignatari2010}
{Pignatari}, M., {Gallino}, R., {Heil}, M., {et~al.} 2010, \apj, 710, 1557

\bibitem[{{Placco} {et~al.}(2014){Placco}, {Frebel}, {Beers}, \&
  {Stancliffe}}]{Placco2014}
{Placco}, V.~M., {Frebel}, A., {Beers}, T.~C., \& {Stancliffe}, R.~J. 2014,
  \apj, 797, 21

\bibitem[{{Prudil} {et~al.}(2021){Prudil}, {Hanke}, {Lemasle}, {Crestani},
  {Braga}, {Fabrizio}, {Koch-Hansen}, {Bono}, {Grebel}, {Matsunaga}, {Marengo},
  {da Silva}, {Dall'Ora}, {Mart{\'\i}nez-V{\'a}zquez}, {Altavilla}, {Lala},
  {Chaboyer}, {Ferraro}, {Fiorentino}, {Gilligan}, {Nonino}, \&
  {Th{\'e}venin}}]{Prudil2021}
{Prudil}, Z., {Hanke}, M., {Lemasle}, B., {et~al.} 2021, \aap, 648, A78

\bibitem[{{Ram{\'{\i}}rez} \& {Mel{\'e}ndez}(2005)}]{RamirezMelendez2005}
{Ram{\'{\i}}rez}, I. \& {Mel{\'e}ndez}, J. 2005, \apj, 626, 465

\bibitem[{{Raso} {et~al.}(2020){Raso}, {Libralato}, {Bellini}, {Ferraro},
  {Lanzoni}, {Cadelano}, {Pallanca}, {Dalessandro}, {Piotto}, {Anderson}, \&
  {Sohn}}]{Raso2020}
{Raso}, S., {Libralato}, M., {Bellini}, A., {et~al.} 2020, \apj, 895, 15

\bibitem[{{Reichert} {et~al.}(2020){Reichert}, {Hansen}, {Hanke},
  {Sk{\'u}lad{\'o}ttir}, {Arcones}, \& {Grebel}}]{Reichert2020}
{Reichert}, M., {Hansen}, C.~J., {Hanke}, M., {et~al.} 2020, \aap, 641, A127

\bibitem[{{Reichert} {et~al.}(2021{\natexlab{a}}){Reichert}, {Hansen}, \&
  {Arcones}}]{Reichert2021}
{Reichert}, M., {Hansen}, C.~J., \& {Arcones}, A. 2021{\natexlab{a}}, \apj,
  912, 157

\bibitem[{{Reichert} {et~al.}(2021{\natexlab{b}}){Reichert}, {Obergaulinger},
  {Eichler}, {Aloy}, \& {Arcones}}]{Reichert2021MRSN}
{Reichert}, M., {Obergaulinger}, M., {Eichler}, M., {Aloy}, M.~{\'A}., \&
  {Arcones}, A. 2021{\natexlab{b}}, \mnras, 501, 5733

\bibitem[{{Roederer} \& {Sneden}(2011)}]{Roederer2011}
{Roederer}, I.~U. \& {Sneden}, C. 2011, \aj, 142, 22

\bibitem[{{Roederer} {et~al.}(2014){Roederer}, {Preston}, {Thompson},
  {Shectman}, {Sneden}, {Burley}, \& {Kelson}}]{Roederer2014}
{Roederer}, I.~U., {Preston}, G.~W., {Thompson}, I.~B., {et~al.} 2014, \aj,
  147, 136

\bibitem[{{Roederer} \& {Thompson}(2015)}]{Roederer2015}
{Roederer}, I.~U. \& {Thompson}, I.~B. 2015, \mnras, 449, 3889

\bibitem[{{Roederer} {et~al.}(2018){Roederer}, {Hattori}, \&
  {Valluri}}]{Roederer2018}
{Roederer}, I.~U., {Hattori}, K., \& {Valluri}, M. 2018, \aj, 156, 179

\bibitem[{{Ruchti} {et~al.}(2013){Ruchti}, {Bergemann}, {Serenelli},
  {Casagrande}, \& {Lind}}]{Ruchti2013}
{Ruchti}, G.~R., {Bergemann}, M., {Serenelli}, A., {Casagrande}, L., \& {Lind},
  K. 2013, \mnras, 429, 126

\bibitem[{{Sanna} {et~al.}(2020){Sanna}, {Franciosini}, {Pancino},
  {Mucciarelli}, {Tsantaki}, {Charbonnel}, {Smiljanic}, {Fu}, {Bragaglia},
  {Lagarde}, {Tautvai{\v{s}}iene}, {Magrini}, {Randich}, {Bensby}, {Korn},
  {Bayo}, {Bergemann}, {Carraro}, \& {Morbidelli}}]{Sanna2020}
{Sanna}, N., {Franciosini}, E., {Pancino}, E., {et~al.} 2020, \aap, 639, L2

\bibitem[{{Sarajedini} \& {Layden}(1995)}]{Sarajedini1995}
{Sarajedini}, A. \& {Layden}, A.~C. 1995, \aj, 109, 1086

\bibitem[{{Schiavon} {et~al.}(2017){Schiavon}, {Zamora}, {Carrera},
  {Lucatello}, {Robin}, {Ness}, {Martell}, {Smith},
  {Garc{\'{\i}}a-Hern{\'a}ndez}, {Manchado}, {Sch{\"o}nrich}, {Bastian},
  {Chiappini}, {Shetrone}, {Mackereth}, {Williams}, {M{\'e}sz{\'a}ros},
  {Allende Prieto}, {Anders}, {Bizyaev}, {Beers}, {Chojnowski}, {Cunha},
  {Epstein}, {Frinchaboy}, {Garc{\'{\i}}a P{\'e}rez}, {Hearty}, {Holtzman},
  {Johnson}, {Kinemuchi}, {Majewski}, {Muna}, {Nidever}, {Nguyen}, {O'Connell},
  {Oravetz}, {Pan}, {Pinsonneault}, {Schneider}, {Schultheis}, {Simmons},
  {Skrutskie}, {Sobeck}, {Wilson}, \& {Zasowski}}]{Schiavon2017}
{Schiavon}, R.~P., {Zamora}, O., {Carrera}, R., {et~al.} 2017, \mnras, 465, 501

\bibitem[{{Schlafly} \& {Finkbeiner}(2011)}]{Schlafly2011}
{Schlafly}, E.~F. \& {Finkbeiner}, D.~P. 2011, \apj, 737, 103

\bibitem[{{Searle} \& {Zinn}(1978)}]{SearleZinn1978}
{Searle}, L. \& {Zinn}, R. 1978, \apj, 225, 357

\bibitem[{{Shetrone} \& {Keane}(2000)}]{Shetrone2000}
{Shetrone}, M.~D. \& {Keane}, M.~J. 2000, \aj, 119, 840

\bibitem[{{Shi} {et~al.}(2018){Shi}, {Yan}, {Zhou}, \& {Zhao}}]{Shi2018}
{Shi}, J.~R., {Yan}, H.~L., {Zhou}, Z.~M., \& {Zhao}, G. 2018, \apj, 862, 71

\bibitem[{{Shibata} \& {Hotokezaka}(2019)}]{Shibata2019}
{Shibata}, M. \& {Hotokezaka}, K. 2019, Annual Review of Nuclear and Particle
  Science, 69, 41

\bibitem[{{Shipp} {et~al.}(2018){Shipp}, {Drlica-Wagner}, {Balbinot},
  {Ferguson}, {Erkal}, {Li}, {Bechtol}, {Belokurov}, {Buncher}, {Carollo},
  {Carrasco Kind}, {Kuehn}, {Marshall}, {Pace}, {Rykoff}, {Sevilla-Noarbe},
  {Sheldon}, {Strigari}, {Vivas}, {Yanny}, {Zenteno}, {Abbott}, {Abdalla},
  {Allam}, {Avila}, {Bertin}, {Brooks}, {Burke}, {Carretero}, {Castander},
  {Cawthon}, {Crocce}, {Cunha}, {D'Andrea}, {da Costa}, {Davis}, {De Vicente},
  {Desai}, {Diehl}, {Doel}, {Evrard}, {Flaugher}, {Fosalba}, {Frieman},
  {Garc{\'{\i}}a-Bellido}, {Gaztanaga}, {Gerdes}, {Gruen}, {Gruendl},
  {Gschwend}, {Gutierrez}, {Hartley}, {Honscheid}, {Hoyle}, {James}, {Johnson},
  {Krause}, {Kuropatkin}, {Lahav}, {Lin}, {Maia}, {March}, {Martini},
  {Menanteau}, {Miller}, {Miquel}, {Nichol}, {Plazas}, {Romer}, {Sako},
  {Sanchez}, {Santiago}, {Scarpine}, {Schindler}, {Schubnell}, {Smith},
  {Smith}, {Sobreira}, {Suchyta}, {Swanson}, {Tarle}, {Thomas}, {Tucker},
  {Walker}, {Wechsler}, \& {DES Collaboration}}]{Shipp2018}
{Shipp}, N., {Drlica-Wagner}, A., {Balbinot}, E., {et~al.} 2018, \apj, 862, 114

\bibitem[{{Sneden}(1973)}]{Sneden1973}
{Sneden}, C.~A. 1973, PhD thesis, The University of Texas at Austin.

\bibitem[{{Sneden} {et~al.}(1997){Sneden}, {Kraft}, {Shetrone}, {Smith},
  {Langer}, \& {Prosser}}]{Sneden1997}
{Sneden}, C., {Kraft}, R.~P., {Shetrone}, M.~D., {et~al.} 1997, \aj, 114, 1964

\bibitem[{{Sneden} {et~al.}(2003){Sneden}, {Cowan}, {Lawler}, {Ivans},
  {Burles}, {Beers}, {Primas}, {Hill}, {Truran}, {Fuller}, {Pfeiffer}, \&
  {Kratz}}]{Sneden2003}
{Sneden}, C., {Cowan}, J.~J., {Lawler}, J.~E., {et~al.} 2003, \apj, 591, 936

\bibitem[{{Sobeck} {et~al.}(2006){Sobeck}, {Ivans}, {Simmerer}, {Sneden},
  {Hoeflich}, {Fulbright}, \& {Kraft}}]{Sobeck2006}
{Sobeck}, J.~S., {Ivans}, I.~I., {Simmerer}, J.~A., {et~al.} 2006, \aj, 131,
  2949

\bibitem[{{Spite}(1992)}]{Spite1992}
{Spite}, M. 1992, in The Stellar Populations of Galaxies, ed. B.~{Barbuy} \&
  A.~{Renzini}, Vol. 149, 123

\bibitem[{{Stromgren} {et~al.}(1982){Stromgren}, {Gustafsson}, \&
  {Olsen}}]{Stromgren1982}
{Stromgren}, B., {Gustafsson}, B., \& {Olsen}, E.~H. 1982, \pasp, 94, 5

\bibitem[{{Tolstoy} {et~al.}(2009){Tolstoy}, {Hill}, \& {Tosi}}]{Tolstoy2009}
{Tolstoy}, E., {Hill}, V., \& {Tosi}, M. 2009, \araa, 47, 371

\bibitem[{{Travaglio} {et~al.}(2004){Travaglio}, {Gallino}, {Arnone}, {Cowan},
  {Jordan}, \& {Sneden}}]{Travaglio2004}
{Travaglio}, C., {Gallino}, R., {Arnone}, E., {et~al.} 2004, \apj, 601, 864

\bibitem[{{Watson} {et~al.}(2019){Watson}, {Hansen}, {Selsing}, {Koch},
  {Malesani}, {Andersen}, {Fynbo}, {Arcones}, {Bauswein}, {Covino}, {Grado},
  {Heintz}, {Hunt}, {Kouveliotou}, {Leloudas}, {Levan}, {Mazzali}, \&
  {Pian}}]{Watson2019}
{Watson}, D., {Hansen}, C.~J., {Selsing}, J., {et~al.} 2019, \nat, 574, 497

\bibitem[{{Webb} \& {Leigh}(2015)}]{Webb2015}
{Webb}, J.~J. \& {Leigh}, N.~W.~C. 2015, \mnras, 453, 3278

\bibitem[{{Westin} {et~al.}(2000){Westin}, {Sneden}, {Gustafsson}, \&
  {Cowan}}]{Westin2000}
{Westin}, J., {Sneden}, C., {Gustafsson}, B., \& {Cowan}, J.~J. 2000, \apj,
  530, 783

\bibitem[{{Woosley} \& {Hoffman}(1992)}]{Woosley1992}
{Woosley}, S.~E. \& {Hoffman}, R.~D. 1992, \apj, 395, 202

\bibitem[{{Worley} {et~al.}(2013){Worley}, {Hill}, {Sobeck}, \&
  {Carretta}}]{Worley2013}
{Worley}, C.~C., {Hill}, V., {Sobeck}, J., \& {Carretta}, E. 2013, \aap, 553,
  A47

\bibitem[{{Yong} {et~al.}(2005){Yong}, {Grundahl}, {Nissen}, {Jensen}, \&
  {Lambert}}]{Yong2005}
{Yong}, D., {Grundahl}, F., {Nissen}, P.~E., {Jensen}, H.~R., \& {Lambert},
  D.~L. 2005, \aap, 438, 875

\bibitem[{{Yong} {et~al.}(2006){Yong}, {Aoki}, {Lambert}, \&
  {Paulson}}]{Yong2006}
{Yong}, D., {Aoki}, W., {Lambert}, D.~L., \& {Paulson}, D.~B. 2006, \apj, 639,
  918

\bibitem[{{Yong} {et~al.}(2008{\natexlab{a}}){Yong}, {Karakas}, {Lambert},
  {Chieffi}, \& {Limongi}}]{Yong2008a}
{Yong}, D., {Karakas}, A.~I., {Lambert}, D.~L., {Chieffi}, A., \& {Limongi}, M.
  2008{\natexlab{a}}, \apj, 689, 1031

\bibitem[{{Yong} {et~al.}(2008{\natexlab{b}}){Yong}, {Lambert}, {Paulson}, \&
  {Carney}}]{Yong2008b}
{Yong}, D., {Lambert}, D.~L., {Paulson}, D.~B., \& {Carney}, B.~W.
  2008{\natexlab{b}}, \apj, 673, 854

\bibitem[{{Yong} {et~al.}(2014){Yong}, {Roederer}, {Grundahl}, {Da Costa},
  {Karakas}, {Norris}, {Aoki}, {Fishlock}, {Marino}, \& {Milone}}]{Yong2014}
{Yong}, D., {Roederer}, I.~U., {Grundahl}, F., {et~al.} 2014, \mnras, 441, 3396

\bibitem[{{Zinn} \& {West}(1984)}]{ZinnWest1984}
{Zinn}, R. \& {West}, M.~J. 1984, \apjs, 55, 45

\end{thebibliography}
\appendix
\section{Systematic errors}
\begin{table*}[htb]
\caption{Systematic error analysis.}
\centering
\begin{tabular}{ccccccc}
\hline\hline       
 & $T_{\rm eff}$ & log\,$g$& [M/H] &  $\xi$ &  &  \\
\rb{Species} & $\pm$60 K & $\pm$0.15 dex & $\pm$0.1 dex &  $\pm$0.1 km\,s$^{-1}$ & \rb{ODF} & \rb{Sys.} \\
\hline       
  & $\pm$0.01 & $\pm$0.07 & $\pm$0.04 & $\mp$0.01 & $-$0.12 &  0.09 \\
\rb{ O\,{\sc  i}} & $\pm$0.01 & $\pm$0.06 & $\pm$0.04 & $\mp$0.02 & $-$0.11 &  0.08 \\
 & $\pm$0.05 & $\mp$0.01 & $\mp$0.01 & $\mp$0.03 & $+$0.02 &  0.06 \\
\rb{Na\,{\sc  i}} & $\pm$0.05 & $\mp$0.01 & $\mp$0.01 & $\mp$0.04 & $+$0.02 &  0.07 \\
 & $\pm$0.04 & $\mp$0.01 & $\pm$0.00 & $\mp$0.06 & $-$0.02 &  0.07 \\
\rb{Mg\,{\sc  i}} & $\pm$0.05 & $\mp$0.01 & $\mp$0.00 & $\mp$0.07 & $-$0.00 &  0.09 \\
 & $\pm$0.04 & $\mp$0.00 & $\mp$0.00 & $\mp$0.01 & $+$0.01 &  0.04 \\
\rb{Al\,{\sc  i}} & $\pm$0.05 & $\mp$0.00 & $\mp$0.01 & $\mp$0.01 & $+$0.02 &  0.05 \\
 & $\mp$0.03 & $\pm$0.04 & $\pm$0.02 & $\mp$0.02 & $-$0.05 &  0.06 \\
\rb{Si\,{\sc  i}} & $\mp$0.01 & $\pm$0.02 & $\pm$0.01 & $\mp$0.02 & $-$0.03 &  0.03 \\
 & $\pm$0.11 & $\mp$0.00 & $\mp$0.01 & $\mp$0.12 & $-$0.06 &  0.16 \\
\rb{ K\,{\sc  i}} & $\pm$0.09 & $\mp$0.00 & $\mp$0.02 & $\mp$0.12 & $-$0.02 &  0.15 \\
 & $\pm$0.07 & $\mp$0.01 & $\mp$0.01 & $\mp$0.10 & $-$0.00 &  0.12 \\
\rb{Ca\,{\sc  i}} & $\pm$0.07 & $\mp$0.01 & $\mp$0.01 & $\mp$0.08 & $+$0.02 &  0.11 \\
 & $\mp$0.01 & $\pm$0.07 & $\pm$0.03 & $\mp$0.08 & $-$0.11 &  0.11 \\
\rb{Sc\,{\sc ii}} & $\mp$0.01 & $\pm$0.06 & $\pm$0.03 & $\mp$0.07 & $-$0.10 &  0.10 \\
 & $\pm$0.11 & $\mp$0.01 & $\mp$0.01 & $\mp$0.09 & $-$0.01 &  0.14 \\
\rb{Ti\,{\sc  i}} & $\pm$0.11 & $\mp$0.00 & $\mp$0.02 & $\mp$0.06 & $+$0.02 &  0.13 \\
 & $\mp$0.02 & $\pm$0.06 & $\pm$0.03 & $\mp$0.10 & $-$0.10 &  0.12 \\
\rb{Ti\,{\sc ii}} & $\mp$0.01 & $\pm$0.05 & $\pm$0.03 & $\mp$0.10 & $-$0.09 &  0.12 \\
 & $\pm$0.12 & $\mp$0.00 & $\mp$0.01 & $\mp$0.04 & $-$0.01 &  0.13 \\
\rb{ V\,{\sc  i}} & $\pm$0.11 & $\mp$0.00 & $\mp$0.01 & $\mp$0.02 & $+$0.02 &  0.11 \\
 & $\pm$0.10 & $\mp$0.01 & $\mp$0.01 & $\mp$0.09 & $-$0.00 &  0.14 \\
\rb{Cr\,{\sc  i}} & $\pm$0.10 & $\mp$0.01 & $\mp$0.02 & $\mp$0.08 & $+$0.02 &  0.13 \\
 & $\pm$0.10 & $\mp$0.00 & $\mp$0.01 & $\mp$0.05 & $+$0.01 &  0.11 \\
\rb{Mn\,{\sc  i}} & $\pm$0.09 & $\pm$0.01 & $\mp$0.00 & $\mp$0.14 & $-$0.02 &  0.17 \\
 & $\pm$0.04 & $\pm$0.01 & $\pm$0.00 & $\mp$0.08 & $-$0.03 &  0.09 \\
\rb{Fe\,{\sc  i}} & $\pm$0.06 & $\pm$0.00 & $\mp$0.00 & $\mp$0.07 & $+$0.00 &  0.09 \\
 & $\mp$0.12 & $\pm$0.07 & $\pm$0.04 & $\mp$0.06 & $-$0.12 &  0.16 \\
\rb{Fe\,{\sc ii}} & $\mp$0.05 & $\pm$0.06 & $\pm$0.03 & $\mp$0.05 & $-$0.10 &  0.10 \\
 & $\pm$0.05 & $\pm$0.02 & $\pm$0.01 & $\mp$0.04 & $-$0.03 &  0.07 \\
\rb{Co\,{\sc  i}} & $\pm$0.07 & $\pm$0.01 & $\mp$0.00 & $\mp$0.02 & $+$0.00 &  0.07 \\
 & $\pm$0.03 & $\pm$0.03 & $\pm$0.01 & $\mp$0.06 & $-$0.05 &  0.08 \\
\rb{Ni\,{\sc  i}} & $\pm$0.05 & $\pm$0.01 & $\pm$0.00 & $\mp$0.05 & $-$0.01 &  0.07 \\
 & $\pm$0.05 & $\pm$0.02 & $\pm$0.01 & $\mp$0.01 & $-$0.03 & 0.07  \\
\rb{Cu\,{\sc  i}} & $\pm$0.07 & $\pm$0.02 & $\mp$0.01 & $\mp$0.01 & $-$0.01 & 0.07 \\
 & $\mp$0.04 & $\pm$0.05 & $\pm$0.02 & $\mp$0.08 & $-$0.07 &  0.11 \\
\rb{Zn\,{\sc  i}} & $\mp$0.02 & $\pm$0.03 & $\pm$0.02 & $\mp$0.06 & $-$0.06 &  0.07 \\
 & $\pm$0.07 & $\mp$0.01 & $\mp$0.01 & $\mp$0.01 & \phs0.01 & 0.07 \\
\rb{Rb\,{\sc  i}} & $\pm$0.07 & $\mp$0.01 & $\mp$0.01 & $\mp$0.01 & \phs0.02 & 0.07 \\
 & $\pm$0.02 & $\pm$0.02 & $\pm$0.04 & $\mp$0.01 & $-$0.10 & 0.06 \\
\rb{Sr\,{\sc ii}} & $\pm$0.02 & $\pm$0.03 & $\pm$0.04 & $\mp$0.05 & $-$0.11 & 0.08 \\
 & $\mp$0.01 & $\pm$0.06 & $\pm$0.03 & $\mp$0.11 & $-$0.10 &  0.13 \\
\rb{ Y\,{\sc ii}} & $\mp$0.01 & $\pm$0.06 & $\pm$0.03 & $\mp$0.06 & $-$0.10 &  0.09 \\
 & $\pm$0.01 & $\pm$0.05 & $\pm$0.02 & $\mp$0.12 & $-$0.07 &  0.13 \\
\rb{Zr\,{\sc ii}} & $\pm$0.01 & $\pm$0.05 & $\pm$0.02 & $\mp$0.07 & $-$0.08 &  0.09 \\
 & $\pm$0.02 & $\pm$0.06 & $\pm$0.03 & $\mp$0.14 & $-$0.12 & 0.16 \\
\rb{Ba\,{\sc ii}} & $\pm$0.02 & $\pm$0.05 & $\pm$0.04 & $\mp$0.12 & $-$0.10 & 0.14 \\
 & $\pm$0.01 & $\pm$0.07 & $\pm$0.04 & $\mp$0.01 & $-$0.11 &  0.09 \\
\rb{La\,{\sc ii}} & $\pm$0.01 & $\pm$0.06 & $\pm$0.04 & $\mp$0.01 & $-$0.11 &  0.08 \\
 & $\pm$0.01 & $\pm$0.06 & $\pm$0.03 & $\mp$0.02 & $-$0.10 &  0.08 \\
\rb{Ce\,{\sc ii}} & $\pm$0.01 & $\pm$0.06 & $\pm$0.03 & $\mp$0.01 & $-$0.09 &  0.07 \\
 & $\pm$0.01 & $\pm$0.06 & $\pm$0.03 & $\mp$0.05 & $-$0.10 &  0.09 \\
\rb{Nd\,{\sc ii}} & $\pm$0.01 & $\pm$0.06 & $\pm$0.03 & $\mp$0.04 & $-$0.10 &  0.08 \\
 & $\pm$0.02 & $\pm$0.06 & $\pm$0.03 & $\mp$0.05 & $-$0.10 &  0.09 \\
\rb{Sm\,{\sc ii}} & $\pm$0.02 & $\pm$0.06 & $\pm$0.03 & $\mp$0.04 & $-$0.09 &  0.08 \\
 & $\mp$0.01 & $\pm$0.07 & $\pm$0.04 & $\mp$0.03 & $-$0.12 &  0.09 \\
\rb{Eu\,{\sc ii}} & $\mp$0.01 & $\pm$0.06 & $\pm$0.03 & $\mp$0.01 & $-$0.10 & 0.07 \\
\hline                  
\end{tabular}
\tablefoot{For each element, the respective top row refers to star 35, while the second one each is for star 46.}
\end{table*}
\end{document}